\documentclass[aps,pre,onecolumn,noshowpacs,superscriptaddress,preprintnumbers,nofootinbib,showpacs,floatfix,12pt]{revtex4-2}%groupedaddress,
\usepackage{epsfig,dsfont,amssymb,amsmath,amsthm,amsfonts,amsbsy,mathrsfs}
\usepackage{lipsum}
\usepackage{color}
\usepackage{graphicx}
\usepackage{subfigure}
\usepackage{float}
\usepackage{xr}
\usepackage{hyperref}
\usepackage{verbatim}
\usepackage{textcomp}
\usepackage{soul}
\usepackage{ulem}
\usepackage{multirow}
\setstcolor{red}

\renewcommand{\thesubsection}{\Alph{subsection}}
\makeatletter
\newcommand*{\addFileDependency}[1]{% argument=file name and extension
  \typeout{(#1)}
  \@addtofilelist{#1}
  \IfFileExists{#1}{}{\typeout{No file #1.}}
}
\makeatother

\begin{document}
%\begin{multicols}{2}

%\title{Stochastic gene regulation dynamics for {\em E.coli} flagellar synthesis in single cells}

%\title{Multiplicative noise underlies Taylor's law in protein concentration fluctuations in growing and dividing cells: Taylor's law and its mechanistic origin}

\title{Multiplicative noise underlies Taylor's law in protein concentration fluctuations in single cells}

\author{Alberto Stefano Sassi}
 \thanks{Equal contributions}
\affiliation{IBM T.J. Watson Research Center, Yorktown Heights, NY 10598, U.S.A}
\author{Mayra Garcia-Alcala}
\thanks{Equal contributions}
\affiliation{Department of Molecular and Cellular Biology, John A. Paulson School of Engineering and Applied Sciences, Harvard University, Cambridge, MA 02138, USA}
\affiliation{Instituto de Ciencias Físicas, Universidad Nacional Autónoma de México, Cuernavaca, Morelos 62210, México}
\author{Philippe Cluzel}
\thanks{Corresponding authors: cluzel@mcb.harvard.edu, yuhai@us.ibm.com}
\affiliation{Department of Molecular and Cellular Biology, John A. Paulson School of Engineering and Applied Sciences, Harvard University, Cambridge, MA 02138, USA}
\thanks{Corresponding authors: cluzel@mcb.harvard.edu, yuhai@us.ibm.com}
\author{Yuhai Tu}
\thanks{Corresponding authors: cluzel@mcb.harvard.edu, yuhai@us.ibm.com}
\affiliation{IBM T.J. Watson Research Center, Yorktown Heights, NY 10598, U.S.A}

\begin{abstract}
\vspace{0.5cm}
    Protein concentration in a living cell fluctuates over time due to noise in growth and division processes. From extensive single-cell experiments by using {\it E. coli} strains with different promoter strength (over two orders of magnitude) and under different nutrient conditions, we found that the variance of protein concentration fluctuations follows a robust square power-law dependence on its mean, which belongs to a general phenomenon called Taylor's law. To understand the mechanistic origin of this observation, we use a minimal mechanistic model to describe the stochastic growth and division processes in a single cell with a feedback mechanism for regulating cell division. The model reproduces the observed Taylor's law. The predicted protein concentration distributions agree quantitatively with those measured in experiments for different nutrient conditions and a wide range of promoter strength. By using a mean-field approximation, we derived a single Langevin equation for protein concentration with multiplicative noise, which can be solved exactly to prove the square Taylor's law and to obtain an analytical solution for the protein concentration distribution function that agrees with experiments. Analysis of experiments by using our model showed that noise in production rates dominates over noise from cell division in contributing to protein concentration fluctuations. 
    %Our model is also used to study cell size homeostasis. We found that the added length between cell division is independent of cell length and the division time depends inversely on the cell length, both in agreement with experiments. 
    %Deviations from the square Taylor's law were found among individual cells in a homogeneous population, which can be explained by including a correlation between production rate and the corresponding noise strength in our model. 
    In general, a multiplicative noise in the underlying stochastic dynamics may be responsible for Taylor's law in other systems.
    %for maintaining cell size homeostasis.  noise sources    The statistics of the protein concentration inside the cell is strongly dependent on the dynamics of cell growth. This is due to the fact that the rate of protein production and cell elongation present significant independent fluctuations, even though their average values are correlated. In this work, we show that the concentration of protein inside the cell,  measured for several cell generations, obey the Taylor's law, which is a power law dependence between the variance and the mean of a stochastic variable. We then propose a mechanistic model of cell growth and division that reproduces the distribution of protein concentration inside the cell and explain why the Taylor's law is followed in this system. The model is based on the coupling between a production variable, the number of ribosomes in the cell, and a division variable, the number of division proteins, that together guarantee the presence of steady state time traces for cell size and protein number.
\end{abstract}
\maketitle
\newpage
\section{Introduction}
Protein concentrations inside a single cell determine functions and behaviors  of the cell~\cite{Cross2007, Eldar2010, Lin2018, Brenner2015, Cao2020}. Given the small size of a cell, dynamics of protein concentration is highly stochastic and there is a significant cell-to-cell variability~\cite{Susman, Wallden, Jun_Curr_Bio, Tanouchi, Tu2020}. %While there are various theoretical models aimed at explaining the statistics of protein copy number in growing and dividing cells, the behavior of the protein concentration has not been addressed to a similar extent.
%{\color{red} \sout{ There are various phenomenological models aimed at describing the fluctuations of the protein content of growing cells}}\cite{BrennerPRE, Bertaux}. 
%{\color{blue}However, there is significantly less abundance in models that are instead \textit{mechanistic}, in the sense that all their components have a clear biological meaning. }
%mechanistic models {\color{blue}describing the interplay between cell growth and protein concentration during each cell cycle.}
%{\color{red}\sout{that take the influence of cell growth and division on the fluctuations in the concentration into account explicitly}}. 
%{\color{red}\sout{The dynamics of each cell cycle has a crucial influence on the protein concentration,}}
Protein concentration dynamics in a cell is determined by gene expression and protein synthesis processes as well as the growth and division processes of the cell, all of which can introduce significant noise~\cite{Paulsson2011}. These different sources of noise make it challenging to study statistics and dynamics of single cell protein concentration.  %As a consequence, cell population measurements provide limited information for studying statistical fluctuations because the cell-to-cell variability and the different stages of a cell cycle are averaged out. 
Recently, advances in single-cell experimental techniques % such as the "mother machine", 
have started to shed lights on this difficult problem~\cite{Cluzel, Jun_Curr_Bio, Weissman2006, Soifer}. 
One of the most revealing single cell measurements is done by using the so called ``mother machine'' where bacterial cells are constrained to grow and divide inside a microfluidic device \cite{Cluzel, Wang2010, Ollion2019}, which allows simultaneous monitoring of the protein copy number as well as the cell length for many cell generations in a controllable and stable environment. % with one of its end exposed to constant flow of fresh media so that a steady chemostatic growth can be ensured for many cell generations. 
%Single-cell techniques allowed improvements in our understanding of mechanisms both concerning the dynamics of cell division \cite{Jun_Curr_Bio, Amir2014, Scott2010} and the statistical properties of the protein content in growing cells \cite{Susman, Brenner2015, BrennerPRE, Bertaux}. {\color{red}\sout{Since they allow to monitor the cell size during full cycles of growth and division for several generations, they have been extensively used to study the characteristics of cell size control.}}

%Among the most remarkable results, there is now a general consensus that bacterial cells increase their size by a constant amount during each cycle \cite{Cross2007, Jun_Curr_Bio, Lin2018, Ho2018, Zheng2016, Huang2017}, independently on the size at birth. This mechanism is described in the so called \textit{ adder} model \cite{Sompayrac1973, Amir2014}, which is supported by several experiments in bacteria and yeasts. Another relevant finding that highlights the importance of single-cell measurements is the deviation from the growth law.  When the rate of growth is tuned by changing the nutrient conditions, in several organisms the population experiments showed that the average size depends on the average growth rate through an exponential dependence \cite{Scott2010, Salman2012, Zheng2016}. However, at the single-cell level the same relationship no longer holds, and cells with the same growth rate do not have an average size that is determined by the same relationship \cite{Jun_Curr_Bio, Lagomarsino2016}.

In this work, we studied the distribution of protein concentration in growing and dividing {\it E. coli} cells by using the mother machine microfluidic device. By measuring the single-cell protein concentration dynamics in a set of {\it E. coli} strains with different promoter strength, we found that the variance of the single cell protein concentration fluctuations is proportional to the square of the mean for all strains over a wide range (over $10^2$ fold change) of the mean protein concentration. Furthermore, we found that the scaled protein concentration distribution follows a universal function independent of the specific value of the mean. %the mean protein concentration.
A power law scaling relationship between variance and mean is generally called the Taylor's law, which is named after Lionel Taylor who first observed it in ecology~\cite{Taylor} and later found to exist in diverse fields from number theory to epidemiology~\cite{Joel_Cohen1, Joel_Cohen2, Keeling, Cohen2013}.
The protein copy number fluctuations were studied in {\it E. coli} by Salman et al~\cite{Salman2012}, where a square dependence of the mean of the protein number with respect to the variance was first reported. 
%This general result was also derived for the more specific system of protein number fluctuations in growing and  dividing cells by Salman et al~\cite{Salman2012}. 
By using a phenomenological model~\cite{Brenner2015, BrennerPRE}, Brenner et al showed that the protein number $N$ in a growing and dividing cell follows approximately a log-normal distribution with a constant variance under the assumption that $N$ increases exponentially with respect to time. The only protein number scale comes from the mean, which is fixed in the phenomenological model as a mathematical constraint. This result is a particular case of a previous more general work on population growth in a  stochastic environment by Lewontin and D. Cohen~\cite{Lewontin1969}, which was later linked to the Taylor's law by J. Cohen et al.~\cite{Cohen_Schuster2013, Cohen2014}. 

However, while previous models~\cite{Brenner2015, BrennerPRE} were able to explain the Taylor's law in protein number fluctuations phenomenologically, it can not be used to explain fluctuations of protein concentration, which depend on fluctuations in both protein number and cell size. Since gene expression and cell size have correlated yet different dynamics during growth and division, the challenge for explaining the observed Taylor's law in protein concentration fluctuations can only be met by understanding both the protein number and cell size dynamics consistently in a growing and dividing cell.

%In this paper,  Brenner2015, BrennerPRE, Susman} and demonstrated analytically using a  . Within the assumptions of exponential growth, it was demonstrated that the protein number follows a log-normal distribution. This behavior is a direct consequence of the fact that the protein number is decreased by a multiplicative factor upon division. %On the other hand, the statistical properties of the concentration, instead of the protein number, remain to be discussed. 
%However, this property is no longer valid in the case of the protein concentration, therefore the distribution of the concentration can only be obtained from an analysis of the stochastic fluctuations around the mean value. What is the reason for the emergence of such a universal scaling relation in the statistics of the protein concentration in growing cells? Is it possible to relate the constants of the Taylor's law relation to meaningful biological parameters?

%The fact that the protein number inside the cell follows the Taylor's law was previously shown in experiments \cite{Salman2012, Susman, Brenner2015, BrennerPRE} and demonstrated analytically using a phenomenological model \cite{Brenner2015, BrennerPRE}. Within the assumptions of exponential growth, it can be shown that the protein number follows a log-normal distribution. On the other hand, the statistical properties of the concentration $c$, instead of the protein number $N$, remain to be discussed. 
In a single cell, dynamics of the protein number (for a particular protein) and the cell size are correlated since they are controlled by the same ribosome-based protein production machinery as well as the same cell division events. However, their dynamics are not identical. For the growth process, since cell size growth involves production of a combination of different proteins/biomolecules, the instantaneous cell size growth rate will be different from that for a particular protein.  When a cell divides, the partition factor (the fraction that goes into the daughter cell) is different for a particular protein than for the cell size. These differences are easy to appreciate as there would be no protein concentration fluctuations if their dynamics were identical. % if their dynamics were perfectly correlated. 

To understand the origin of the observed Taylor's law in protein concentration fluctuations, we developed a simple unified model to describe dynamics of both protein number and cell size during growth and division. In the model, the growth of both protein number and cell size depend on the number of a common production machinery (complex) with production (growth) rates different for protein number and cell size. The production machinery, which can be characterized by the number of ribosomes ($R$) in the cell, has its own dynamics, which is controlled by the same production and division processes as for a specific protein. Instead of enforcing a mean protein number as an {\it ad hoc} mathematical constraint as in previous models for protein number fluctuations~\cite{BrennerPRE} and cell size regulation~\cite{Amir2014}, a division control variable (molecule $Z$) is introduced in our model. $Z$ also follows the same production and division dynamics as that of a protein (or $R$). The probability of cell division increases sharply when the number of molecule $Z$ crosses a certain threshold $Z_0$~\cite{Fantes1975, Wold1994, Boye2003, Donachie2003, Basan}. In this paper, we use this minimal mechanistic model to study protein concentration fluctuations and compare the theoretical results quantitatively with single cell data from the mother machine experiments.    %When the cell divides, the partition error, which is defined as the difference of the partition for protein number and cell size are considered to be independent. 

\section{Results}
\subsection{Protein concentration fluctuations follow Taylor's law}
\begin{figure}[ht]
\begin{center}
\includegraphics[clip, width=1.0\linewidth]{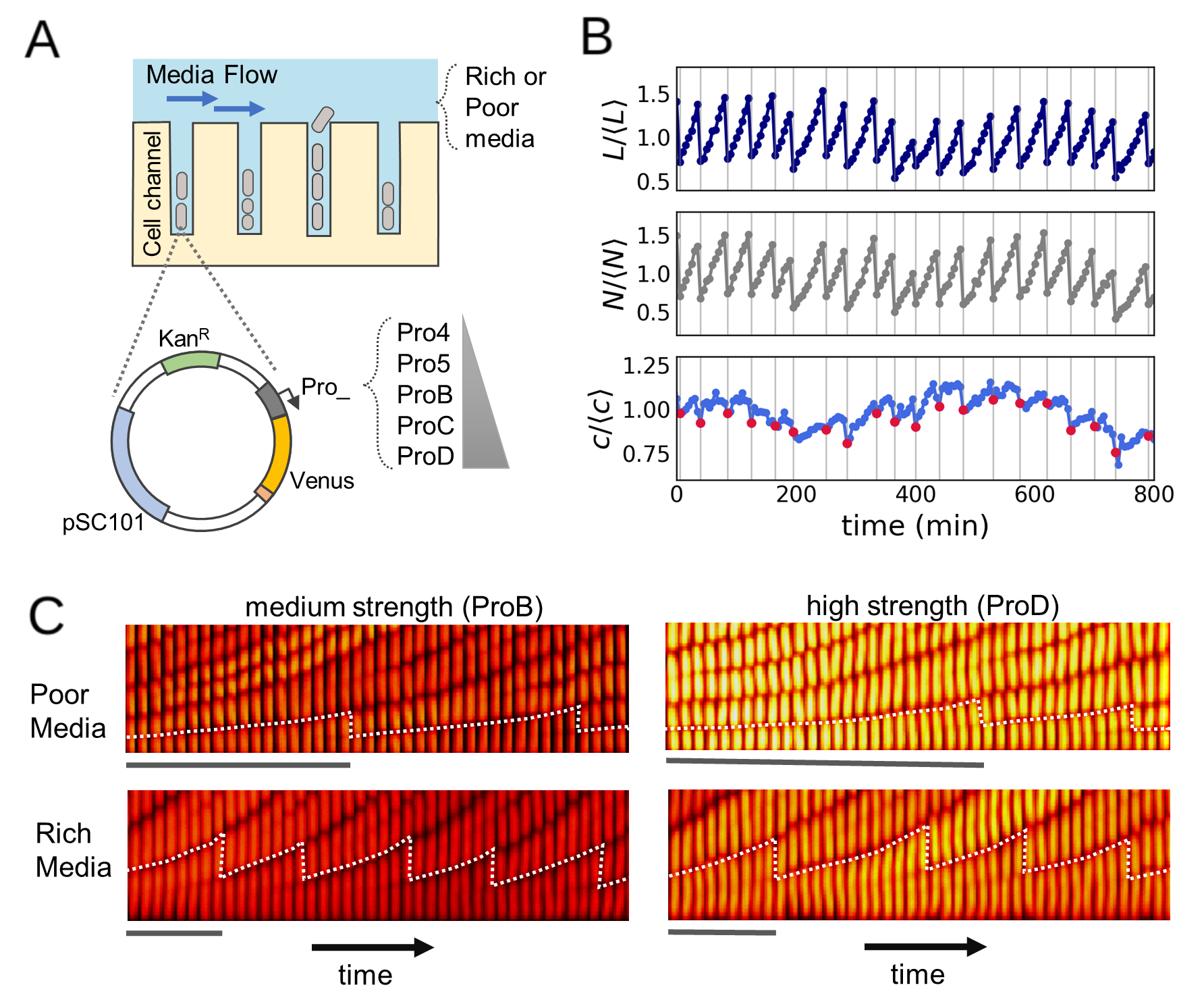}%{experimental_traces.pdf}
\caption{\label{fig:time_traces}
(A) Schematics of the experimental setup. Using the microfluidic device \textit{mother machine} and fluorescent microscopy, we monitored individual cells of various {\it E. coli} strains growing under a constant flow of either rich or poor media.  Each strain contains a plasmid that codifies for the fluorescent protein Venus under the control of an artificial promoter from a set of promoters with increasing strengths, Pro\{4,5,B,C,D\}~\cite{Davis2011}.
(B) Typical time traces of a single cell containing the ProB promoter and growing in rich medium: normalized cell length $L$ (top), total fluorescence $N$ (middle), and fluorescence density $c$ (bottom). 
The vertical lines and the red dots in the bottom plot indicate the time of a cell  division.
(C) Kymographs illustrating the fluorescence level of strains with the promoters ProB (left) and ProD (right) controlling Venus expression, growing in poor (top) and rich (bottom) media. Each frame is separated from the following one by a time interval of 5 minutes. The bars below the kymographs show the span of a cell trace from birth until division.  }
\end{center}
\end{figure}

We constructed a set of \textit{E. coli} strains that produce the fluorescent protein Venus % at various levels,
(Fig.~\ref{fig:time_traces}A). Each strain contains a plasmid that expresses the Venus gene controlled by a promoter from a set of constitutive promoters with different strengths (see Appendix~\ref{sec:experiments} for details about the experiments)~\cite{Davis2011}. Using the mother machine technique \cite{Wang2010, Cluzel, Jun_Curr_Bio, Tu2020}, and fluorescent microscopy, we monitored individual cells of such strains under steady-state conditions, with a constant flow of either rich or poor media. % (see Materials and Methods). 
% COMMENT (MAYRA): I think the following is better for either the caption of Fig.1  or Methods and Materials. 
%For each combination of nutrient conditions and promoter strength, we measured the cell size and the fluorescence for $\sim 20$ different mother cells and $\sim 50$ cell generations per mother cell. 
For each combination of nutrient conditions and promoter strength, we measured the cell size and the fluorescent intensity protein for $\sim 20$ different mother cells and $\sim 50$ cell generations per mother cell.
We assume that the intensity of the fluorescence of a cell at time $t$, $N(t)$, is proportional to the number of Venus protein, and thus the fluorescence density (fluorescence divided by cell size, indicated as $c$) can be considered a proxy of the protein concentration. 

In Fig.~\ref{fig:time_traces}B we show a typical time trace of the normalized cell length (size) $L$ (top), total fluorescence $N$ (middle), and fluorescence density $c(\equiv N/L)$ (bottom) for many cell generations. Both protein number (fluorescence intensity) and cell length follow continuous exponential growth that is interrupted by periodic  ``discontinuous" reductions due to cell division. The 
%This is particularly clear when we consider the 
fluorescence density $c$ is much more continuous than $N$ and $L$ and the influence of cell division is not strong. This is easy to understand because the gross effect of cell division for both $N$ and $L$ is canceled out in $c$. However, even though cell division does not affect the protein concentration the same way as the protein number or cell size, we also note from Fig.~\ref{fig:time_traces}B (bottom) that there is a small but observable change in $c$ at cell division times (red dots), which suggests that cell division is likely a source of noise for protein concentration.  %We can see that in the cell size and in the total fluorescence there are fluctuations with time scales that can be  larger than the duration of a cell cycle. 

%By using the mother machine \cite{Wang2010, Cluzel, Jun_Curr_Bio, Tu2020}, \textit{E. coli} cells were kept in the exponential phase for many generations under a given nutrient condition. %so that we could continuously measure the cell size and protein number for many cycles of growth and division. 
%In our experiments, a fluorescent protein is controlled by using different promoters with different strength ({\color{red} Maybe some experimental details here from Philippe or Marya?}). 
%The protein number inside a cell is proportional to the fluorescent intensity that is measured every $5$ minutes. 
%Fluorescence single-cell measurements of the concentration $c$ of a given protein in dividing cells, using the {\em mother machine} technique \cite{Wang2010}, 

\begin{figure}[ht]
\begin{center}
\includegraphics[clip, width=1.0\linewidth]{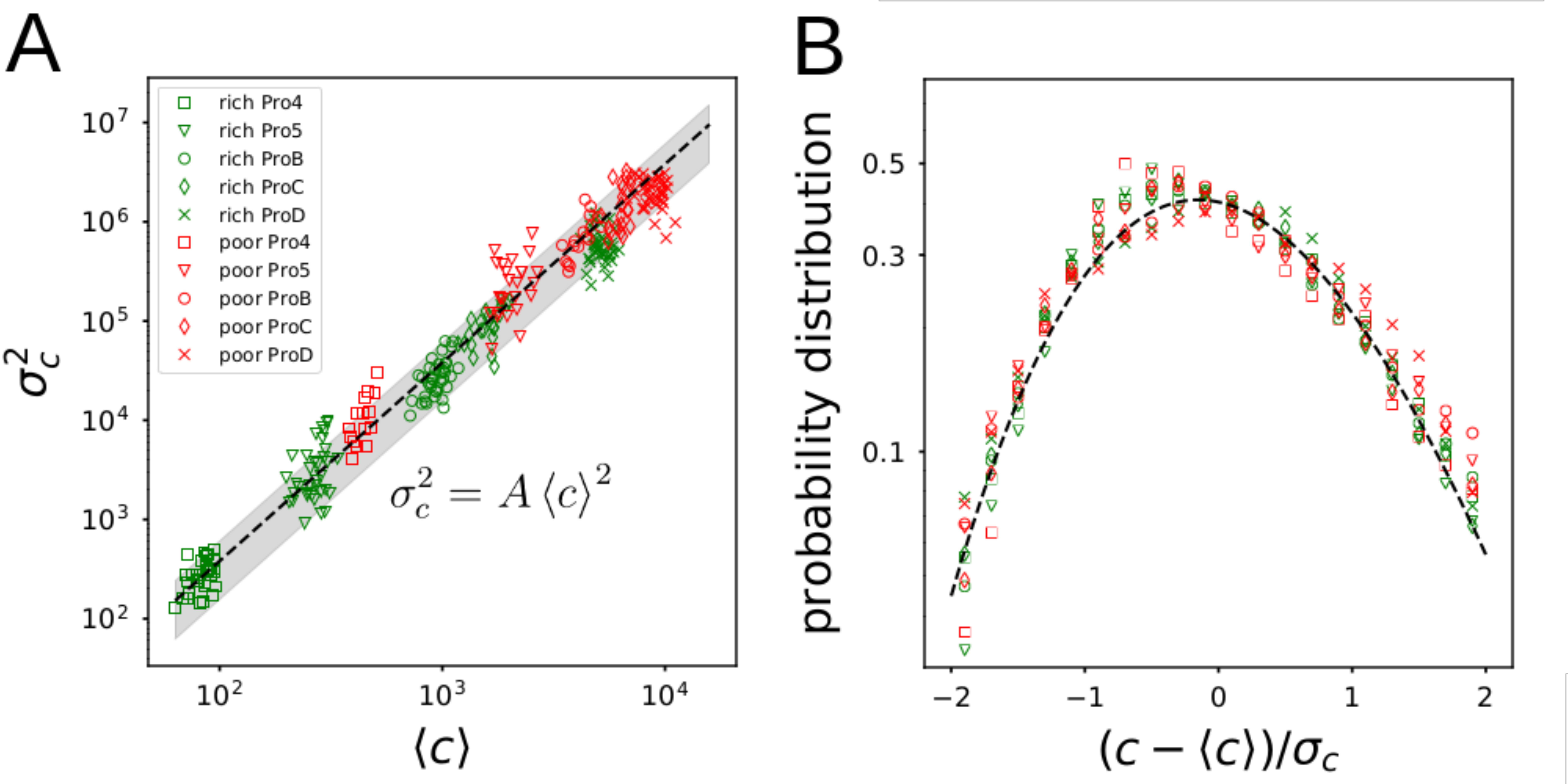}
\caption{\label{fig:distribution_C_exp} 
The Taylor's law and the universal distribution of the normalized protein concentration fluctuations. (A) Variance ($\sigma_c^2$) of the protein concentration versus the mean value ($\langle c\rangle$) for individual cells. The poor and rich nutrient conditions are labeled by red and green colors, respectively. Different symbols correspond to cells with different promoters. All data from different cells with different promoters and growth condition  follow the square Taylor'law: $\sigma_c^2=A\langle c\rangle^2$ with $A=0.038\pm 0.015$. The black dashed line and the gray thick line around it reflect the mean and standard deviation of $A$. (B) Distributions for the normalized fluctuations of $c$ for different nutrient conditions and promoter strengths shown in (A) collapse. The dashed line is obtained from the analytical expression Eq.~\ref{eqn:distribution} from our model.}
\end{center}
\end{figure} 
To study how fluctuations in cell size and protein number depend on the mean growth and transcription rates, we did experiments using several constitutive promoters with different strengths~\cite{Cluzel} and also with two nutrient (growth) conditions~\cite{Jun_Curr_Bio}. Qualitatively, as shown in Fig.~\ref{fig:time_traces}C, a stronger promoter leads to a higher fluorescence intensity; whereas a rich nutrient condition leads to a faster growth and a shorter average division time $\Delta T$.

From each combination of promoter and growth condition, we determined the distribution of the protein concentration, $P(c)$, from measurements such as those shown in Fig.~\ref{fig:time_traces} for many ($\sim 20$) individual mother cells. From the distribution,  we computed the mean $\langle c\rangle$ and the variance $\sigma_c^2$. As shown in Fig.~\ref{fig:distribution_C_exp}A, the variance and the mean follow the square Taylor's law, i.e., the variance $\sigma_c^2$ is proportional to the square of the mean $\left<c\right>$: 
\begin{equation}\label{eqn:Taylor}
    \sigma_c^2 = A\left<c\right>^2\,, 
\end{equation}
with the prefactor constant $A$ within a narrow range $A=0.038\pm 0.015$ for all promoters and nutrient conditions studied in this paper. This is remarkable considering that the power law scaling relation (Eq.~\ref{eqn:Taylor}) is valid for two orders of magnitude of $\left<c\right>$ across different promoter strength and nutrient conditions.  %The multiplicative factor obtained from these data is $A=0.04\pm 0.01$. 
Moreover, the probability distributions of the normalized concentration $(c-\left<c\right>)/\sigma_c$ for all promoters and nutrient conditions collapse onto the same universal curve as shown in Fig.~\ref{fig:distribution_C_exp}B. 
In the following sections we develop a mechanistic model to explain these experimental results. 

%An analogous relation between the mean and the variance was previously measured in the case of the protein number rather than the protein concentration \cite{Salman2012, Susman, Brenner2015, BrennerPRE}. 
%By means of a phenomenological model, it was shown that the  Taylor's law for the protein number is due to the fact that the protein number statistics is well approximated by a log-normal distribution. 
%Indeed, single-cell fluorescent measurements of dividing cells showed that the variance of the protein number (total fluorescence) is proportional to the square of the mean value, and this relation is valid for a large number of different experimental set-up. 

%{\color{red}In the case of protein number, the Taylor's law is a natural consequence of the division process, in which both the protein number and the cell length are roughly halved after each cell cycle \cite{BrennerPRE}. For the concentration, the two effects are on average compensated -since the partition of the length and the protein number occurs simultaneously, and on average it is symmetric in both cases- therefore, to explain the Taylor's law, there is the need to study the stochastic  fluctuations around average values. }
%({\color{blue}This paragraph is not in line with the topic of the rest of the section and it is not a result. Something on this line is probably already said in the introduction so I believe that I should probably simply delete it.})

\subsection{A minimal model for growth and division explains Taylor's law}
\label{sec:B}

Cell growth and division are complex processes involving many regulatory proteins and pathways, which are beyond the scope of this paper. Here, we aim to develop a simple model that captures the most salient features in the underlying molecular mechanism governing protein concentration fluctuations in growing and dividing cells in a mother machine setup. 
%The model must take the cell cycles of growth and division into account explicitly. 
In the minimal model, the growth and division processes in a cell are controlled by a \textit{production} variable $R$ and a \textit{division} variable $Z$, respectively. All dynamic variables in a cell such as the cell length $L$, protein number $N$, as well as $R$ and $Z$ themselves are controlled by $R$ and $Z$.

\begin{figure}[ht]
\begin{center}
\includegraphics[clip, width=0.8\linewidth]{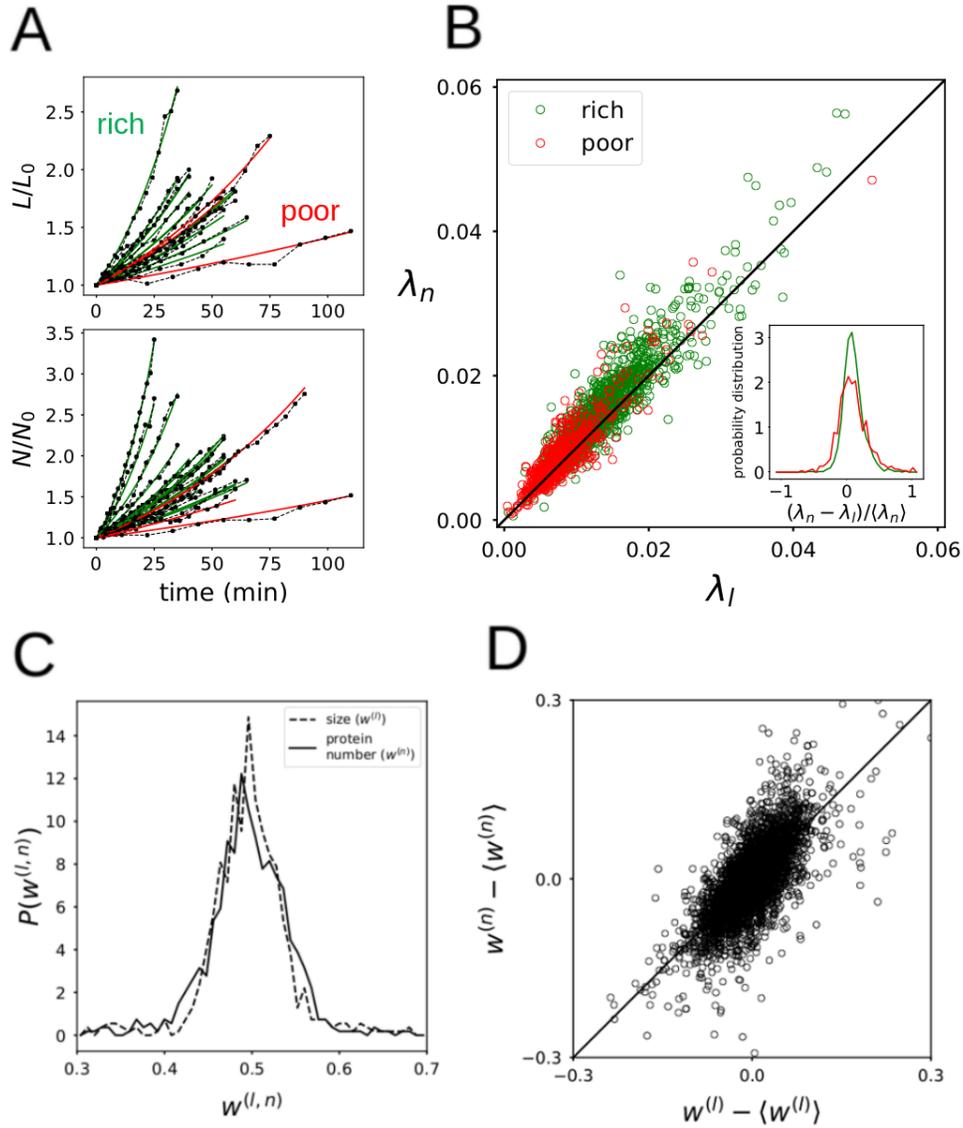}
\caption{\label{fig:growth_rates}Noise sources due to growth and division. (A) The effective growth rates $\lambda_l$ and $\lambda_n$ were obtained by fitting the cell size and protein number dynamics during each growth period (between two consecutive division events) with exponential functions. An example of the fitting is shown for $L$ (top) and $N$ (bottom), different colors correspond to different nutrient conditions. (B) $\lambda_l$ versus $\lambda_n$ shows high but not perfect correlation. Deviations from the diagonal black line correspond to a noise source for $c$ due to growth. The inset shows the distribution of the relative difference between $\lambda_l$ and $\lambda_n$. (C) Distribution of the partition factors  $w^{(l,n)}$ for cell size and protein number. (D) The correlation between $w^{(l)}$ and $w^{(n)}$ is high but not perfect. Deviations from the diagonal black line correspond to another noise source for $c$ due to division.} %Scatter plot of the deviation from the average of the partition variables of cell size and protein number. }
\end{center}
\end{figure}

A common production variable $R$ that is shared by other variables of the system is supported by the fact that growth rates of $L$ and $N$ during cell growth are not independent. As shown in
Fig.~\ref{fig:growth_rates}A, during each growth period between two consecutive divisions, we can fit the cell length and protein number dynamics to exponential functions and determine the growth rates $\lambda_l$ and $\lambda_n$ for $L$ and $N$, respectively. In Fig.~\ref{fig:growth_rates}B, we plot $\lambda_n$ versus $\lambda_l$ for all growth periods and for all three nutrient conditions, which clearly shows that the two growth rates are highly correlated. To account for this strong correlation, all growth rates in our model are assumed to be proportional to the common  production variable $R$, which can be interpreted as the number of active ribosomes in the cell. In previous experiments~\cite{Scott2010}, it was shown that the growth rate depends linearly on the RNA/protein ratio. Since the total RNA content in a cell is a good measure of the ribosome number, these experiments support the assumption that growth rates are proportional to $R$. %made in our model. % suggests that growth rates can be considered as linearly dependent on the ribosome number in the cell. %ribosome number is the main determinant for controlling the protein number growth dynamics and also that translation happens at constant rate for a given ribosome number ({\color{red} Not sure I understand this argument}) ({\color{blue}Here my goal was to justify the choice of the ribosome number as the relevant variable for the growth (and expression) rates. The best argument I found is the fact that in is this article they show that the RNA/protein ratio (proxy for ribosome content?) has a significant influence on the growth rate.}). 

%(A) The effective growth rates $\lambda_l$ $\lambda_n$ was obtained by fitting the cell size with exponential functions. A sample of cell cycles is shown for different nutrient conditions in the case of $L$ (top) and $N$ bottom. (B) Effective rate of cell growth as a function of the effective rate of protein number $N$, for different nutrient conditions. The black line corresponds to $\lambda_l=\lambda_n$, the result expected in the absence of noise. (inset) Distribution of the relative difference between the rates. 
%(A) Distribution of the partition variable $w_i$ for cell size and protein number. (B) Scatter plot of the deviation from the average of the partition variables of cell size and protein number.
%{\color{red}(I put citations here for the moment)}
%{\color{blue}The division variable $Z$ controls the timing of cell division \cite{Basan, Fantes1975, Wold1994, Boye2003, Donachie2003} }. 

In order to maintain cell size homeostasis, a feedback mechanism is needed to control cell division. Indeed, as pointed out in ~\cite{BrennerPRE, Brenner2015}, if cell divisions were to occur at independent random time intervals, even with a fixed mean the accumulation of variation in division times away from their mean would lead to divergence of cell length fluctuation. In bacterial cells, cell division is regulated by the division protein FtsZ~\cite{Si, Dai1991, Mukherjee1993}, which assemble into a ring (the Z ring) localized at the future division site of the cell. The completion of the Z ring, together with other proteins, is critical for cell division. Motivated by these experimental facts and by following previous theoretical work~\cite{ Fantes1975, Wold1994, Boye2003, Donachie2003, Basan}, we assume that the probability for cell division increases sharply with $Z$ when $Z$ crosses certain threshold $Z_0$. Since the Z ring formation depends on oligomerization of FtsZ at a localized site the threshold is set over the total number of proteins as opposed to its concentration.
%it is impossible to directly impose a discontinuity at time instants that are separated by intervals $\Delta T$ (see scheme in Fig.~\ref{fig:division_scheme}) distributed around the observed average value $\tau_d$, and are chosen independently with respect to any other stochastic variable. Indeed, with that approach, the production variable $Q$ used alone would still lead to a divergence in the main relevant quantities over time, because the fluctuations from the average would accumulate \cite{BrennerPRE}. 
%We will show that the feedback mechanism which guarantees the cell size homeostasis is a natural consequence of the coupling between the production and the division variables and there is no need for any additional phenomenological convergence factors. 
%\subsection{Experimental results suggest that the elongation and the size at birth are independent}
%Since the cell size $L$ is on average proportional to the total number of proteins \cite{Si}, its rate must be also proportional to the number of ribosomes. %We indicate as $N$ the number proteins of which we are interested to study the statistics. 

\begin{figure}[ht]
\begin{center}
\includegraphics[clip, width=0.7\linewidth]{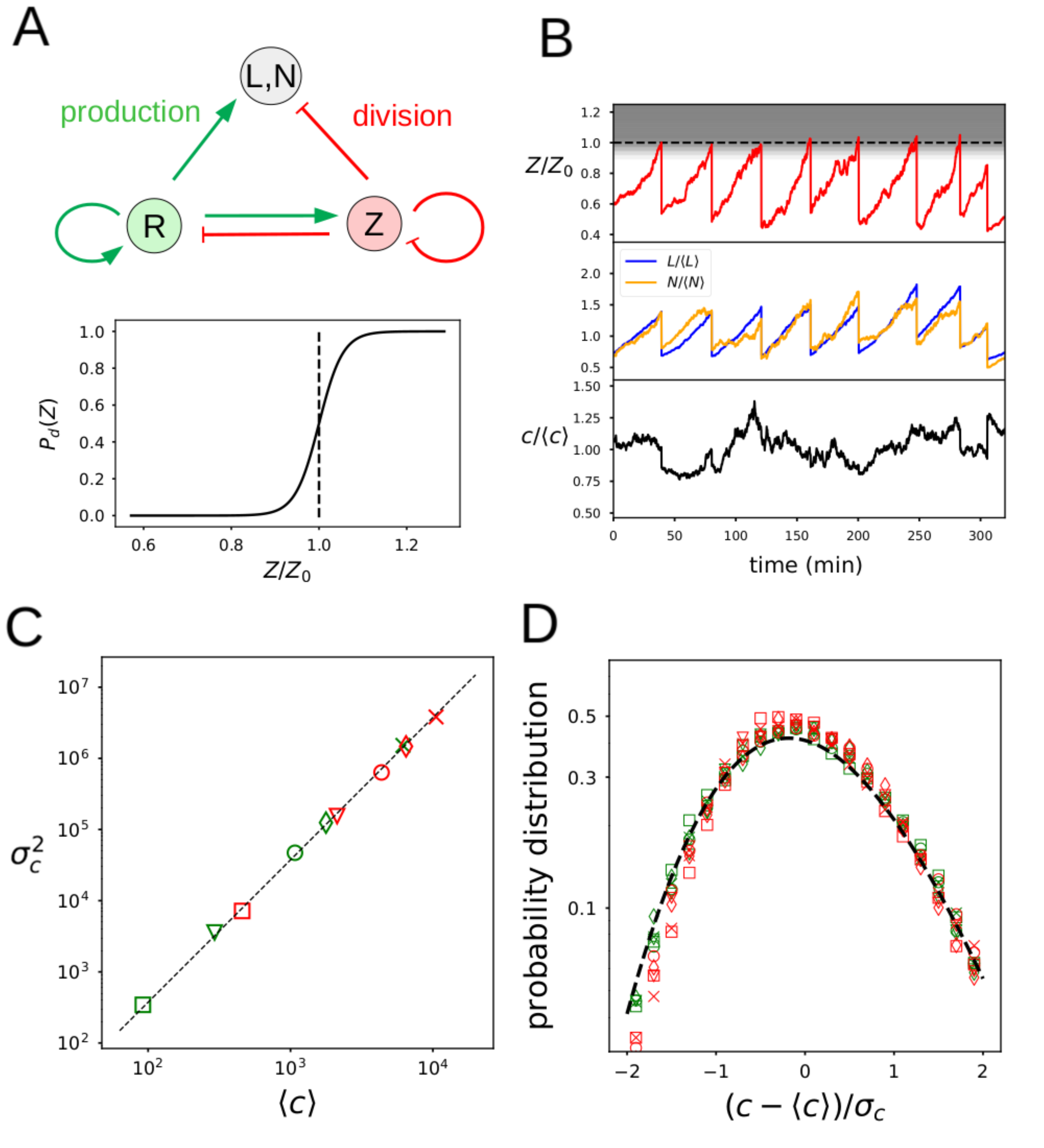}
\caption{\label{fig:traces_XLN}
The minimal mechanistic model and its behaviors. (A) Top: Illustration of the 4-node model. All production (growth) processes (green arrows) depend on $R$. The division process, which reduces all variables (red lines), is controlled by $Z$. Bottom: the probability rate of division $P_d(Z)$ as a function of the division protein number $Z$. (B) Time traces of normalized division protein (red), protein number $N$ (orange), cell size $L$ (blue), and protein concentration $c=N/L$  (black). In the top panel we show in gray the region where the division probability is large (the dotted line is for $Z=Z_0$). (C) Variance of the concentration versus its mean from our model for different promoters (different symbols) and under different nutrient conditions (green: rich medium; red: poor medium). The same symbols and colors are used as experimental data shown in Fig.~\ref{fig:distribution_C_exp}A for easy comparison. %See Appendix~\ref{sec:simulations} for parameters used in the model. 
(D) Distributions of the normalized fluctuations in the concentration obtained with the same parameters as in (C) collapse onto the same curve. The dashed line is from the analytical expression Eq.~\ref{eqn:distribution} obtained from our model. % These two last panels were obtained through a numerical solution of the stochastic equations using different values of $K_n$. 
The parameters for $P_d(Z)$ were $\Delta Z/Z_0=2.7\times 10^{-2}$ and $\Delta t=2\,\mbox{min}$. See Appendix~\ref{sec:simulations} for details of the simulations and parameters used. }
\end{center}
\end{figure}

Taken together, our model is represented schematically in Fig.~\ref{fig:traces_XLN}A (top). There are $4$ extensive variables: the cell size (length) $L$ and the protein number $N$ are ``observable" variables that can be directly measured in the mother machine experiments;  the number of growth complexes (ribosomes) $R$ and the number of division proteins (FtsZ) $Z$ are ``hidden'' variables that control the growth and division of all variables including themselves. The stochastic dynamics of the $4$ variables in the mother cell are described by the following stochastic ordinary differential equations:
\begin{align}
    \frac{dL}{dt}&=R\,K_l(1+\eta_l)-L\sum^{n_d(t)}_{i=1}w^{(l)}_{i}\,\delta(t-t_i(Z)), \label{eqn:main_eq} \\
    \frac{dN}{dt}&=R\,K_n(1+\eta_n)-N\sum^{n_d(t)}_{i=1}w^{(n)}_{i}\,\delta(t-t_i(Z)), \label{eqn:N}\\      
    \frac{dR}{dt}&=R\,K_r(1+\eta_r)-R\sum^{n_d(t)}_{i=1} w^{(r)}_{i}\,\delta(t-t_i(Z)) \\
    \frac{dZ}{dt}&=R\,K_z(1+\eta_z)-Z\sum^{n_d(t)}_{i=1}w^{(z)}_{i}\,\delta(t-t_i(Z))\label{eqn:Q} ,     
\end{align} 
%     \frac{dN}{dt}&=Q\,K_n(1+\eta_n)-N\sum^{n_d(t)}_iw^n_i\,\delta(t-t_i)\nonumber\\
which all have the same general form. The first and second terms on the right side of each equation represent effects of growth and division\footnote{We only consider cell division as the cause of protein number reduction here. However, including protein degradation will not change the main results in this paper due to the multiplicative nature of the degradation process.}, respectively. The parameters $w^{(j)}_i$ ($j=n,l,r,z$) are the partition factors, which is the fraction of variable (protein number) $j$ that goes to the tracked daughter cell upon the $i$-th division event; $t_i(Z)$ is the time of the $i$-th division that depends on $Z$ (see below); $K_j$ ($j=n,l,r,z$) are the mean growth (production) rates; $\eta_j$ ($j=n,l,r,z$) are the fractional noise in growth rates, which are taken to be Gaussian white noise here with zero mean and variance $\Delta_j$; and $n_d(t)$ is the number of divisions up to a time $t$: $ n_d(t)=\int_0^t\, dt'\sum^{\infty}_i \delta(t'-t_i(Z))$. 

The feedback control for cell division is modeled here by a soft-thresholding process with a simple logistic function (Fig.~\ref{fig:traces_XLN}A (bottom)):
\begin{equation}\label{eqn:probability_rate}
    P_d(Z)=\frac{\Delta t^{-1}}{1+\exp\left[(Z_0-Z)/\Delta Z\right]}
\end{equation}
which is characterized by three parameters each with clear biological meaning. The threshold value $Z_0$ is the value of $Z$ at which the division probability increases sharply, with the sharpness determined by $\Delta Z\ll Z_0$ ($P_d$ is simply a step function when $\Delta Z=0$).
%We put the threshold on the number of proteins rather than their concentration because the division proteins are known to be strongly localized \cite{Si} ({\color{blue}I have already explained this concept earlier}). 
Once cell division starts, it can take a finite time to complete. This small but finite time scale is given by $\Delta t$ in Eq.~\ref{eqn:probability_rate}. Other forms of $P_d(Z)$ with the same general properties considered here were used without affecting the general results. %Even in the presence of this feedback triggering the division, the value that can be chosen for the production rates are not entirely unconstrained. Indeed, we must be sure that after $X$ has crossed the threshold $X_0$ the cell divides fast enough, so that $X$ does not become more than twice the threshold, otherwise the switch would be  left permanently \textit{on} and there would not be a steady state. 
The choice of parameters used in our simulations is discussed in Appendix~\ref{sec:simulations}.

There are two sources of noise due to fluctuations in growth rates and in partition factors, respectively. The noise in growth rates ($\eta_{n,l}$) for $N$ and $L$ can be estimated from measurements and analysis shown in Fig.~\ref{fig:growth_rates}A\&B. Similarly, the noise in partition factors ($w^{(n,l)}$) can be determined from the measured distributions for $w^{(n,l)}$ as shown in Fig.~\ref{fig:growth_rates}C. Though there is a strong correlation between these two partition factors (Fig.~\ref{fig:growth_rates}D), their difference remains significant and it gives rise to another source of noise for $c$.
%{\color{red} I am not sure what this following argument is trying to say. Also, for Fig. 3D, why don't we just plot $w^{(n)}$ versus $w^{(l)}$ the same as in Fig.3B?} ({\color{blue}the initial purpose was to show that the fluctuations upon division are not negligible when compared to the fluctuations during growth, but it is indeed a bit confused and I'm ok with removing this paragraph altogether.})   This is evident when we compare the noise in the growth with the noise in the division. We evaluated the ratio between the cell size in two consecutive time frames $t_i$ and $t_{i+1}$ during growth, and the ratio has mean $\left<L(t_{i+1})/L(t_i)\right>=1.08$ 
%and standard deviation $\sigma\left[L(t_{i+1})/L(t_i)\right]=0.032$, so that the coefficient of variation was $\phi_{g}=0.03$. On the other hand, when the two time frames were before and after a division, the ratio has mean $\left<L_{b}/L_{a}\right>=1.97$ and standard deviation $\sigma\left[L_{b}/L_{a}\right]=0.22$, with a coefficient of variation $\phi_{d}=0.11$.
%This can be seen when we evaluate the growth noise used in the simulation to reproduce the experiments  $\left<q\right>K_l\sqrt{\Delta_l}\simeq 1.3*10^{-2}$ and the division noise, that can be obtained directly from the single-cell experimental data, $\sqrt{\Delta_w/\left<\Delta T\right>}\simeq 1.5*10^{-2}$. 

%For the protein content, we add to the system a stochastic variable, $N$, whose average production rate is proportional to $Q$ like for any other protein, and that is also halved upon division:
% \begin{equation}\label{eqn:N}
%    \frac{dN}{dt}=Q\,K_n(1+\eta_n)-N\sum^{n_d(t)}_iw^{(n)}_i\,\delta(t-t_i)\,.     
% \end{equation}

We studied our model by solving the stochastic equations (Eqs.~\ref{eqn:main_eq}-\ref{eqn:Q}) numerically with physiologically reasonable parameters, some of which are estimated from the mother machine experiments. In Fig.~\ref{fig:traces_XLN}B, we show the time traces of different variables as well as the dynamics of the protein concentration. The cell division control by the variable $Z$ can be seen from the top panel in Fig.~\ref{fig:traces_XLN}B which shows that $Z$ follows the same growth and division dynamics as other variables and division occurs with a high probability as $Z$ (red) crosses over the threshold $Z_0$ (dashed line).  As shown in the middle panel in Fig.~\ref{fig:traces_XLN}B, both $L$ (blue) and $N$ (orange) follows the same general growth and division pattern as that of $Z$, but their dynamics are not identical. As a result, the protein concentration $c$ (black) shows smaller but finite and continuous fluctuations that are different from those in $N$ or $L$. The behaviors of $N$, $L$, and $c$ from our model resemble closely with those from experiments shown in Fig.~\ref{fig:time_traces}. 

We also studied cell size homeostasis during growth~\cite{Jun_Curr_Bio, Huang2017}, i.e., the dependence of elongation $\Delta L$ and division time $\Delta T$ on initial cell size $L_0$ during each cell cycle under different growth conditions. The model results are in quantitative agreement with experiments, which validates the model for describing cell growth and division (see Appendix~\ref{sec:homeostasis} for details). 

To quantitatively compare the results from the model with the experimental results, we tune the rates ($K_{n,l,r,z}$) and the noise strength ($\Delta_{n,l,r,z}$), in accordance with experimental data for different promoters and in different nutrient conditions. Given that the promoter strength primarily influences the expression rate of the corresponding proteins, we assumed that the change in promoter is captured in the model by a change in the value of $K_n$, so that the rate of increase in $N$ would be directly affected. The mechanism by which nutrient condition influences the kinetic rates of growth and expression is beyond the scope of this work. Here, we treat the nutrient dependence within our model phenomenologically based on experiments. In particular, from the experiments, the average division time $\left<\Delta T\right>$ is longer in the poor nutrient condition, whereas the average length size $\left<L\right>$ is smaller. We find that the simplest way to account for this observed difference in our model is by assuming that the production rate for the division variable $Z$, $K_z$, is roughly independent of nutrient conditions while the other two rates for growth-related variables ($R$ and $L$) are scaled by a common nutrient-dependent constant $\xi$, i.e., $K_i\rightarrow \xi K_i$ with $i=r,l$ with $\xi=1$ for rich medium and $\xi =0.6$ for poor medium (see Appendix~\ref{sec:homeostasis} for details). Operationally, we first tune $K_l$ and $K_r$ to match the observed cell growth statistics in different nutrient condition before we tune $K_n$ to properly fit the value of $\left<c\right>$. See Appendix~\ref{sec:simulations} for details of the simulations and parameter choice. 

%To test Taylor's law, we change the mean growth/production rate $K_n$ to mimic the different constitutive promoters used in our mother machine experiments with all the other parameters left unchanged.
%We solved the stochastic equations numerically and  the mean concentration $\langle c\rangle$ over 2 orders of magnitude. 
As shown in Fig.~\ref{fig:traces_XLN}C, where the variance of $c$ is plotted as a function of its mean, the simulation results from our model obey the same square Taylor's law as in the experiments with the coefficient of proportionality $A= 0.034\pm 0.006$ that is in quantitative agreement with the experiments. Furthermore, the normalized distribution function for $c$ shows the same collapse for all values of $K_n$ as shown in Fig.~\ref{fig:traces_XLN}D. 

%{\color{red} In a new paragraph, we need to discuss about how we incorporate the n. c. by changing the rates, maybe taking some material from the SI.}  %used (Fig.~\ref{fig:traces_XLN}D). 

%{\color{blue}It's time to discuss the addition of $N$.} We ran the simulations for different value of $K_n$, which is the parameter which is considered to be proportional to the promoter strength. As shown in Fig.~\ref{fig:histo_c_model}, we obtain the proper dependency of the variance of the concentration from the mean value, and also the same collapse for the distributions of the normalized fluctuations of $c$.
\subsection{Multiplicative noise underlies Taylor's law in protein concentration fluctuations}
In the previous section, we studied protein concentration fluctuations by simulating the full model (Eqs.~\ref{eqn:N}-\ref{eqn:Q}) numerically. Here, we show the emergence of Taylor's law by using a mean field approximation, which leads to a closed form relation between the variance and the mean of $c$ and an analytical expression for the probability distribution of $c$. We start by considering the equation for $c$, obtained from Eqs.~\ref{eqn:main_eq}\&\ref{eqn:N} (see  Appendix~\ref{sec:Langevin_c} for details):
\begin{equation}\label{eqn:conc}
   \frac{dc}{dt}=\hat{K}_n-\hat{K}_l\,c+c\sum_i^{n_d(t)}\left[\left(w_i^{(l)}-w_i^{(n)}\right)\delta(t-t_i)\right]\,,
\end{equation}
% \frac{dc}{dt}=q\left(K_n(1-\eta_n)+c\,K_l(1+\eta_l)\right)+c\sum_i^{n_d(t)}\left[\left(w_i^{(l)}-w_i^{(n)}\right)\delta(t-t_i)\right]\,,     
%\frac{dc}{dt}=\hat{K}_l\left(\mu(1+\chi_n)-c\,(1+\chi_l)\right)+c\sum_i^{n_d(t)}\left[\left(w_i^{(l)}-w_i^{(n)}\right)\delta(t-t_i)\right]
%where we have defined the concentration of ribosomes $q=\frac{Q}{L}$.
where we have defined the effective rates $\hat{K}_l=rK_l(1+\eta_l)$ and $\hat{K}_n=rK_n(1+\eta_n)$, with $r=R/L$. The difference $\delta w_i=w_i^{(l)}-w_i^{(n)}$ has zero mean $\left<\delta w_i\right>=0$ and a delta-function correlation $\left<\delta w_i\delta w_j\right>=2\Delta_w\delta_{ij}$ %with $\Delta_w$ the noise strength %of the partition difference. 
%is a stochastic variable with zero mean, which is also delta-correlated since subsequent divisions are independent. 
%, and there is no correlation between the partitions of subsequent divisions. 
as $\delta w_i$ for different division events can be considered independent. For a time scale longer than the average division time $\langle \Delta T\rangle$, we can define an average (mean-field) cell division noise $\eta_d$ whose correlation has the form
%\begin{equation}
   $ \left<\eta_d(t_1)\eta_d(t_2)\right>=2\Delta_d\delta(t_1-t_2)$
%\end{equation}
with $\Delta_d=\Delta_w/\left<\Delta T\right>$ the cell division noise strength. 
The variables $\hat{K}_l$ and $\hat{K}_n$ in Eq.~\ref{eqn:conc}
can be written as: $
    \hat{K}_{l,n}=\left<\hat{K}_{l,n}\right>(1+\chi_{l,n})$
where $\chi_{l,n}=\frac{\delta r}{\langle r\rangle }+\eta_{n,l}$ is the fractional noise with $\delta r(=r-\langle r\rangle)$ and $\langle r\rangle $ the fluctuation and the mean of $r$. %and $\chi_n$ are stochastic variables with zero mean that contain contributions from $r$ and $\eta_{l,n}$.
Thus, by taking the mean-field approximation in the long time limit, the Langevin equation (Eq.~\ref{eqn:conc}) can be re-written as:
\begin{equation}\label{eqn:c_with_mu}
    \frac{dc}{dt}=\left[\left<\hat{K}_l\right>(\mu-c)+\mu\eta_a+c\eta_m\right]\,,
\end{equation}
where $\mu=\left<\hat{K}_n\right>/\left<\hat{K}_l\right>$ is the average of $c$: $\langle c\rangle =\mu$, which is varied experimentally by changing the promoter strength or the nutrient condition, whereas  $\eta_a=\left<\hat{K}_l\right>\chi_n$ and $\eta_m=\eta_d-\left<\hat{K}_l\right>\chi_l$ are the two noise terms for $c$. 
%The strength of the additive noise is 
%\begin{equation}
%    \Delta_a= K_l^2(\left<\delta q^2\right>\tau_r+\left<q\right>^2\Delta_n)\,,
%\end{equation}
%where $\tau_r$ is a time scale of the system, whereas the strength of the multiplicative noise is 
%\begin{equation}
 %   \Delta_m = \Delta_d + K_l^2(\left<q\right>^2 \Delta_l+\left<\delta q^2\right>\tau_r)\,.
%\end{equation}

It is interesting to note that both noise terms ($\eta_a$ and $\eta_m$) in Eq.~\ref{eqn:c_with_mu} are multiplied by either the mean concentration ($\mu$) or the instantaneous concentration ($c$) itself. As a result of the multiplicative nature of the noise terms, Eq.~\ref{eqn:c_with_mu} is invariant if $c$ and $\mu$ are scaled by an arbitrary constant factor\footnote{This scale invariance is rooted in the invariance of the full model (Eqs.~\ref{eqn:N}-\ref{eqn:Q}) under the transformation: $N\rightarrow \lambda N$, $K_n \rightarrow \lambda K_n$ for arbitrary positive $\lambda$.}. This scale invariance immediately suggests that the distribution of $c/\mu$ is independent of $\mu$ and the variance is proportional to the square of the mean. Indeed, by solving the Fokker-Planck equation corresponding to Eq.~\ref{eqn:c_with_mu}, we derive an exact relationship between the variance and the mean (see Appendix~\ref{Sec:TL_derivation} for details):  
\begin{equation}\label{eqn:TL}
    \sigma^2_c = A \left<c\right>^2\,,
\end{equation}
and the prefactor $A$ can be determined analytically: 
\begin{equation}
A=\frac{\Delta_g+\Delta_d}{\left<\hat{K}_l\right>-\Delta_m}\,,
\label{eq:prefactor}
\end{equation}
where $\Delta_g = \left<\hat{K}_l\right>^2(\Delta_{n}+\Delta_{l})$ is the strength of the growth-dependent noise due to fluctuations in growth and production rates for cell size ($\Delta_l$) and protein number ($\Delta_n$);  $\Delta_m=\Delta_d+\left<\hat{K}_l\right>^2(\Delta_{(r)}+\Delta_l)$ is the strength of the noise $\eta_m$ with $\Delta_{(r)}$ the noise strength for $\delta r/\langle r\rangle $.

Eq.~\ref{eqn:TL} is a direct confirmation of the Taylor's law, with the same exponent $p=2$ as observed in experiments. Moreover, the analytical expression for the prefactor constant $A$ (Eq.~\ref{eq:prefactor}) reveals explicitly that both the growth rate noise ($\Delta_g$) and the cell-division partition noise ($\Delta _d$) contribute to $A$. These noise strength ($\Delta_g$ and $\Delta_d$) can be determined quantitatively. For example, in our model (Eqs.~\ref{eqn:main_eq}-\ref{eqn:Q}) with a typical set of parameters that fit the averaged experimental results shown in Fig.~\ref{fig:traces_XLN}, the noise strengths from division and growth can be obtained: $\Delta_d=3.7\times 10^{-5}\,\mbox{min}^{-1}$, $\Delta_g=3.5\times 10^{-4}\, \mbox{min}^{-1}$, which shows that the contribution from growth rate fluctuations is $\sim 10$-fold stronger than that from cell division.  Analysis of different noise sources for experiments with different strains under different growth conditions confirms that the noise from growth rate fluctuation is in general stronger than that from the partition error during cell division  (see Table~\ref{table:parameters} in Appendix~\ref{Sec:noise_strength}). Our analysis also shows that the average value of $A$, which is a measure of overall noise, is slightly smaller in rich medium than that in poor medium. However, the difference in the average $A$ between the two nutrient conditions is relatively small and is comparable with the difference in $A$ among individual cells with the same promoter and under the same nutrient condition (See Fig.~\ref{fig:A_bar_chart} in Appendix~\ref{Sec:noise_strength} for a detailed comparison).   % analysis. % the  that the noise strength as reflected in the value for can be found in the SI. %Combining modeling and  which shows that protein concentration fluctuations are dominated by noise in the growth rate although the contribution from noise in division can not be ignored.  

Finally, by solving the steady state Fokker-Planck equation for Eq.~\ref{eqn:c_with_mu}, we obtain an analytical expression for the distribution function of $c(\ge 0)$: 
\begin{equation}\label{eqn:distribution}
    P(c)= \frac{1}{Z}\left(\frac{\mu^2\Delta_a+2\mu\Delta_{am}c+\Delta_mc^2}{\left<\hat{K}_l\right>}\right)^{-1-\frac{\left<\hat{K}_l\right>}{2\Delta_m}}\exp\left[\frac{2\left<\hat{K}_l\right>(\Delta_{am}+\Delta_m)\tan^{-1}\left[\frac{\mu\Delta_{am}+\Delta_m c}{\mu\rho_\Delta}\right]}{2\Delta_m\rho_\Delta}\right]\,,
\end{equation}
and $P(c<0)=0$ where $Z$ is the normalization constant, $\Delta_a =\left<\hat{K}_l\right>^2(\Delta_{(r)}+\Delta_n)$ is the noise strength for $\eta_a$, $\Delta_{am}=-\left<\hat{K}_l\right>^2\Delta_{(r)}$ is the correlation between $\eta_a$ and $\eta_m$,
%, $\Delta_m$ and $\Delta_a$ are the noise strengths of $\eta_m$ and $\eta_a$ respectively, $\Delta_{am}$ is the correlation of the two noises
and $\rho_\Delta=\sqrt{\Delta_a\Delta_m-\Delta_{am}^2}$. %Details of the derivation for Eq.~\ref{eqn:distribution} can be found in {\color{blue}Sec.~\ref{Sec:}}. 
There is a negligibly small value for $P(c)$ at $c=0$, which is caused by assuming $\eta_a$ to be an unbounded Gaussian noise. For large $c\gg \mu$, $P(c)$ decays as a power law $\sim c^{-2(1+\frac{\left<\hat{K}_l\right>}{2\Delta_m})}$, which is different from a log-normal distribution but similar to an inverse Gamma distribution, which is the solution for $P(c)$ in the limit of negligible addition noise ($\Delta_a=\Delta_{am}=\rho_{\Delta}=0$). %  which is the distribution that we would obtain in the case the multiplicative noise is dominant with respect to the additive noise. On the other hand, for negative $c$ the distribution can have small yet finite values. This non-physiological effect is due to the contributions from the additive noise, but with the parameters inferred from the fluorescence experiments $c$ the probability that $c<0$ is negligible. 
In Figs.~\ref{fig:distribution_C_exp}B\&\ref{fig:traces_XLN}D, we plotted the analytical distribution function given in Eq.~\ref{eqn:distribution} (black dashed line), which quantitatively agrees with experimental results and simulation results from the full stochastic model. %The analytical distribution function quantitatively agrees with the collapsed protein concentration distributions obtained from experimental data for different strains with different promoter strength. 

\section{Discussion}
%{\color{blue} Discussion topics include: 1) Taylor's law -- ubiquitous scaling law in nature without mechanistic understanding ... our work is one of the few cases where microscopic origin of the Taylor's law is reveal .... The approach (Langevin equation with proper noise consideration) may be generally applicable to understand Taylor's law in other systems....  2) Cell division dynamics -- our model provide a starting minimal mechanistic model for studying noise and fluctuation during the cell growth and division process ... many directions to pursue, e.g., partition of resources, connection to the growth law, the nature of molecule X, etc... 3) Cell-cell variability -- relate to other cell-cell variability studies (e.g., for cell growth law, etc.) Discuss our novel approach of looking at cell-cell variability as a way to probe hidden correlations between different stochastic variables, and how this approach may be generally applicable for studying other cell-cell variability problems. what else do we want to say here? 4) More topics? }  

In this paper, we developed a minimal mechanistic model to study stochastic dynamics of protein concentration in single cells over a long timescale that spans many generations. The balance between growth and division processes is key to maintain a dynamic equilibrium for cells. Here, we showed that the stochastic growth and division processes in a single cell are the two dominant noise sources contributing to  fluctuations in protein concentration. By considering these two stochastic processes explicitly and consistently in our model, we were able to obtain the square variance-mean scaling relation (Taylor's law) as well as the universal distribution for the normalized concentration, both of which are in quantitative agreement with our mother machine experiments with different promoters and under different nutrient conditions. We have used our model to analyze previous experiments~\cite{Salman2012} under different experimental conditions and obtained consistent results although the data there cover a much smaller range ($\sim 1/2$ decade) of mean protein concentrations (see Fig.~\ref{fig:Salman_distribution} in Appendix for details). 

There are two central regulatory variables in our model, $R$ and $Z$, which control growth and division, respectively. Both $R$ and $Z$ can be considered as large complexes with multiple components, however, each of them has identifiable key components: $R$ is associated with ribosome, whereas $Z$ is associated with FtsZ. In our minimal model, we only implemented the simplest possible interactions between $R$ and $Z$ -- the growth rates of all proteins (including $R$ and $Z$ themselves) are proportional to $R$; and the probability of cell division increases sharply when $Z$ crosses a certain threshold $Z_0$. Despite its simplicity, the minimal model is able to reproduce the observed statistics of protein concentration fluctuations and cell size homeostasis. More importantly, the minimal model provides a mathematical framework to ask further questions regarding molecular origins of the growth and division processes. For the division control, one key question is about the molecular origin for the cell division regulation, which is modeled here by the assumption (hypothesis) that the probability of division increases sharply when $Z$ crosses a threshold $Z_0$. 
%It would be interesting to study whether and how this threshold crossing mechanism is caused by the cooperative FtsZ polymerization process during Z ring formation, which is the keystone for cell division in bacterial cells. 
For the growth process, we did not take into account the way resources are allocated to the production of different protein classes in our model where all production rates are linearly proportional to $R$. The effect of proteome allocation can be studied in our model by allowing different substrates (mRNA's) to compete for the same finite pool of $R$ with different affinities, which results in a nonlinear dependence of the production rates on $R$. It would be interesting to study proteome allocation~\cite{Bertaux} in dividing cells by introducing nonlinear growth rates in  our model.  %taken into account, as it was previously done in other works that were not focusing on the distribution of protein concentration .  

In this paper, we focused on explaining the square Taylor's law (TL) with $p=2$ for protein concentration fluctuations. However, our model also indicates how the exponent $p$ may deviate from $2$. In particular, from the analytical expression for the variance-average dependence (Eq.~\ref{eqn:TL}), it is easy to see that if the prefactor $A$ is correlated with the mean $\mu(\equiv\langle c\rangle)$, then the variance-mean relation would deviate from the square TL, e.g., if $A$ depends on $\mu$ in a power law form, $A\sim \mu ^\alpha$, then the variance-mean relation follows the TL with an exponent $p=2+\alpha$. From the expression for the prefactor $A$ (Eq.~\ref{eq:prefactor}), such a correlation can be introduced by having a correlation between the noise strength ($\Delta_{g,d}$) and $\mu$ and/or a correlation between the time scale $\tau_c\equiv \langle \hat{K}_l\rangle^{-1}$ and $\mu$. In our experiments, the mean concentration is varied by changing the promoter strength characterized by $K_n$ in our model. Since $K_n$ does not appear in the expression for $A$, there is no correlation between $A$ and the mean $\mu$, which leads to the observed square TL. However, $\mu$ also depends on the elongation rate $K_l$, which affects $A$. Therefore, if the mean concentration $\mu$ can be varied by changing the elongation rate $K_l$, our model would predict a deviation from the square TL, which may be tested experimentally.

Taylor's law is an ubiquitous scaling law observed in a plethora of different systems, from the occurrence of measles cases~\cite{Keeling} to the share price fluctuations in  stock market~\cite{Eisler}. However, these observations of Taylor's law remain largely empirical with little understanding of their mechanistic origins. Here, we showed that the square Taylor's law (exponent $p=2$) exists generally in systems whose stochastic dynamics can be described by a Langevin equation with multiplicative noise, i.e., the noise term is multiplied by the variable of interest. Even though we focused on studying protein concentration fluctuations in bacterial cells such as \textit{E. coli}, the general theoretical framework used in this paper may be applicable to other systems where the dominant noise source is multiplicative. Different TL exponents observed in different systems may be caused by possible correlations between key parameters in the system (e.g., the relaxation timescale and the noise strength) and the mean.      % to any system that can be reduced to a similar equation in a mean field approximation.

\section{Acknowledgments}
We thank Dr. Hanna Salman for providing data from ~\cite{Salman2012} and Dr. Joel Cohen for discussions regarding Taylor's law and comments on our manuscript. The work by ASS and YT are supported by a NIH grant (R35GM131734 to YT). The work by MG and PC are supported by a NIH grant (R01GM134275 to PC) and a NSF grant (1615487 to PC).
%\appendix
%\setcounter{section}{0}
%\renewcommand{\thesection}{S\arabic{section}}
%\sectionfont{\MakeUppercase}
\appendix
\renewcommand{\thesubsection}{\Alph{subsection}}
\section*{Appendix}
%\maketitle
\subsection{Experimental Methods}\label{sec:experiments}
%({\color{red}Alberto: in this subsection I removed the bullet points and just left the paragraphs without a subtitle. })
%\begin{itemize}

%\item \textbf{Plasmid construction}
A vector with the artificial Pro5 promoter \cite{Davis2011} upstream the gene of the fluorescent protein Venus followed by the rrnB T1 terminator was inserted into a pSC101 plasmid with the kanamacyn antibiotic resistance gene. It was assembled using the isothermal assembly protocol to construct pPro5Venus plasmid. It was transformed into DH5$\alpha$ strain (New England Biolabs) for selection and amplification. The plasmid sequence was verified by sequencing and named pPro5Venus. 

The rest of the plasmids were constructed out of pPro5Venus. The complete sequence of the plasmid was amplified with primers that helped swap the 10 box with the respective sequence from other promoters from the set published in \cite{Davis2011}. Each plasmid was circularized using the NEB KLD Enzyme Mix (New England Biolabs).

%\item \textbf{Strain construction}
The background strain for this work is MGR-E98K \cite{Cluzel}, which is E. coli MG1655 (CGSC \texttt{\#}6300) with a point mutation in MotA that disables  rotation and prevents cells from swimming out of the cell channels in the microfluidic device. Each  plasmid was transformed into the strain by electroporation, the final set of strains is named ProVenus set. Selected colonies had kanamycin resistance and were fluorescent under the microscope. The content was verified by colony PCR. Using flow cytometry, it was confirmed that the set of strains have different fluorescent levels.

%\item \textbf{Growth Media} 
Two growth media were used to grow the cells in the microfluidic device. They were selected considering that wild-type \textit{E. coli}'s growth rate was significantly higher in one media compared to the other on a batch experiment.
\underline{Rich media}: MOPS EZ rich defined medium from Teknova, supplemented with 0.4\% glycerol.
\underline{Poor media}: M91X, 2 mM MgSO\textsubscript{4}, 0.1 mM CaCl\textsubscript{2}, 0.5\% Casamino acids, 0.4\% Glycerol, 1 ug/ml Thiamine.
In both media, it was added Pluronic F-108 (Sigma-Aldrich) to final concentration of 0.85 g/liter, to act as a surfactant in the microfluidic device.

%\item \textbf{Cell preparation}
Each ProVenus strain was grown overnight in 1 ml of the same growth media that was used in the microfluidic experiment.
In each microfluidic experiment, there were 4 different stains in the same device to ensure they grow in the same conditions avoiding day-to-day variations in the setup. In all experiments the wild-type strain was included and the remaining strains were chosen so that they have distinguishable fluorescence levels.
The overnight cultures were centrifuged at 6000 g for 1 min, and from each of them the same volume of the pellet was taken and mixed by gently pipetting. 

%\item \textbf{Microfluidic experimental setup}
The microfluidic device used in this work is an adaptation of the mother machine described in \cite{Wang2010}. The design of the device and the protocol for the mold construction can be found in \cite{Cluzel}.
For every experiment, a microfluidic device was cast from the same mold.
The cells are loaded into the device by pipetting into it the high-density mixture of strains.
The loaded microfluidic device is connected to a pump that delivers growth media with a constant flow of 5 ul/min. It is placed under the inverted microscope in an incubated box at 30\textdegree C during all the experiment.

%\item \textbf{Fluorescence microscopy}
Cells were tracked using imaging microscopy by taking phase contrast and fluorescent images with the YFP channel every 5 min for ~24 hrs.
The microscope setup was controlled using custom software on MATLAB 2013a  interfacing with $\mu$Manager 1.4. Multiple positions of the device were captured.

%\item 
%\textbf{Cell segmentation and tracking}
We used the software Bacmman for cell segmentation and tracking \cite{Ollion2019}.

%\end{itemize}

%\title{Supplementary Information}
%\section{Supplementary Information}
\subsection{Details of the numerical simulations}\label{sec:simulations}
We solved our stochastic model (Eqs.~\ref{eqn:main_eq}-\ref{eqn:Q} in the main text) numerically. Between two consecutive divisions, the integration of the ODEs was performed with the Euler method. At each time step $\delta t$ a division could take place with a probability $P_d(Z(t))\delta t$, where $Z(t)$ is the value of the division protein number at the time $t$. In case of division, every variable would be multiplied by a factor $f$, where $f=1/2(1+\epsilon_d)$, where $\epsilon_d$ is a Gaussian variable with variance $\sigma_d^2$. We denote $\tau(=45\,\mbox{min})$ as the average division time in rich nutrient conditions, and a small time step is chosen to be $\delta t= 10^{-3}\tau$. The parameters chosen for $P_d(Z)$ were $\Delta Z/Z_0=2.7\times 10^{-2}$ and $\Delta t=2\,\mbox{min}$. The rate of expression of the division protein was $K_z=1\,\mbox{min}^{-1}$.  The noise strengths for $\eta_n$, $\eta_z$, $\eta_l$ and $\eta_r$ were $\Delta_n=1.2\,\mbox{min}$, $\Delta_z=1\,\mbox{min}$, $\Delta_l=0.2\,\mbox{min}$ and $\Delta_r=1$ respectively. %Without loss of generality, the noise in $\eta_r$ could be set to zero, since we were not interested in the time traces of $Q$ specifically and every noise coming from $Q$ could be included in the noises of the other variables. 

The standard deviation for the partition error was $\sigma_d=0.1$, inferred from experiments. We tuned the parameter $K_l$ in such a way that the average size at division was the same as in the experiments and the parameter $K_n$ for the concentration to match the value of the fluorescence traces for every nutrient condition (see Table~\ref{table:Kn_Kl} for the numerical value of $K_l$ and $K_n$). In most cases the simulation were run for a total time $T_t=5\times 10^3\,\mbox{min}$, but in the case we wanted to obtain the probability distribution we used $T_t=5\times 10^4\,\mbox{min}$ to sample a larger statistics.

\begin{table}
\begin{tabular}{ |c|c|c| }
 \hline
%Rich A & 0.506 & 2.4 & 0.017 & 0.59 & 0.0002 & 4.5e-05 & 12 & 0.034 \\
 & Rich & Poor \\
 \hline
\multirow{5}{2em}{$K_n$} & 0.014 & 0.04 \\ 
& 0.04 & 0.2\\
& 0.16 & 0.4\\
& 0.26 & 0.61\\
& 0.92 & 0.97\\
 \hline
\multirow{1}{2em}{$K_l$} & $1.5\times10^{-4}$ & $9\times10^{-5}$ \\ 
 \hline
\end{tabular}
\caption{\label{table:Kn_Kl}Rates $K_n$ and $K_l$ used in the simulations for the two nutrient conditions. The units of measurement are such that $RK_n$ is in units of fluorescence per minute, while $RK_l$ is in $\mu m$ per minute, where $R$ is a variable indicating the number of ribosomes. }
\end{table}

We do not have direct measurements of the ribosome number. However, there are some choices of the  parameters that are constrained by the experimental results.
Indeed, if we consider the equation for $R$, for small noise, the solution between two cell divisions is given by
\begin{equation}\label{eqn:sol_r}
    R(t)=R_0\exp{\left[K_r t\right]}\,.
\end{equation}
Since we want to have $\left<R(t_f)\right>=2\left<R(t_{in})\right>$, where $t_f$ and $t_{in}$ are the final and initial time of a given cell cycle respectively, in order to have a stationary state, the average division time must be
\begin{equation}\label{eqn:ln2}
    \tau_d=\left<t_f-t_i\right>=\ln{2}/K_r\,,
\end{equation}
independently of any other parameter. Therefore, if we set $K_r=\ln{2}/\tau_d$, where $\tau_d=\left<\Delta T\right>$ is taken from the experiments, the simulation will lead to an average division time that is coherent with the results of the experiments performed with the mother machine technique. 

\subsection{The minimal growth-division model explains observed cell division statistics} \label{sec:homeostasis}

There have been extensive experimental study and analysis on the dependence of the elongation $\Delta L$ and the division time $\Delta T$ on the initial cell size $L_0$ during each cell cycle under different growth conditions. Here, we first briefly describe correlations among these quantities from our mother machine experiments and then we check if our model can reproduce the observed results quantitatively.
\begin{figure}[ht]
\begin{center}
\includegraphics[clip, width=0.8\linewidth]{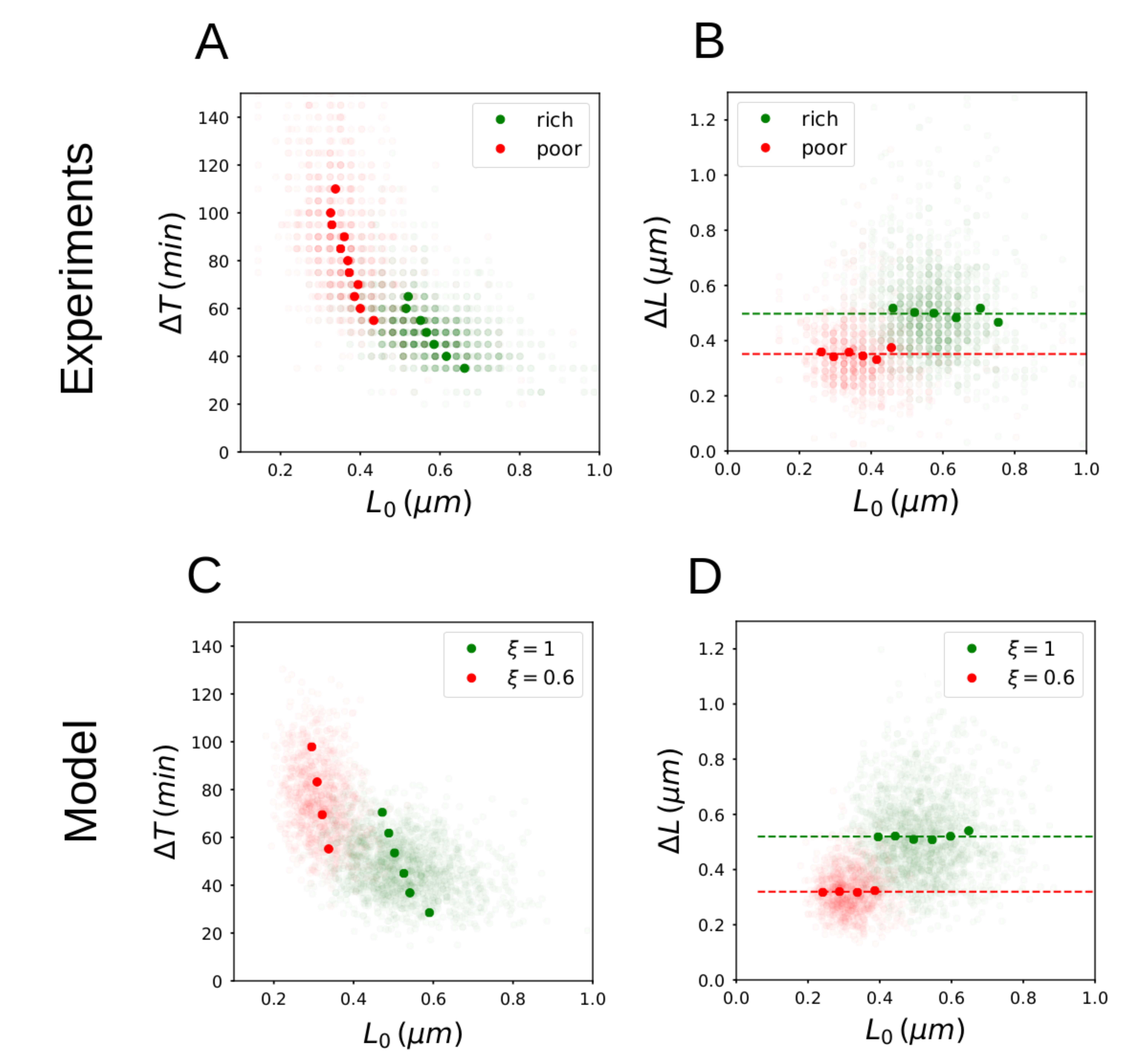}
\caption{\label{fig:exp_vs_model}Comparison between experiments and model results. (A) Division time $\Delta T$ versus the cell size at birth $L_0$ from the mother machine experiment for rich and poor nutrient conditions. Each lighter dot represents a single cell cycle data. Darker dots are bin averages. (B) Elongation $\Delta L$ versus $L_0$ for the same experiments as in (A). (C,D) Same plots as (A) and (B) respectively, from numerical simulations of our model. The same $P_d(Z)$ as in Fig.~\ref{fig:traces_XLN} is used. 
%The noise strengths are $\Delta_z=1\,\mbox{min}$ and $\Delta_l=0.2\, \mbox{min}$, inferred from experimental data. 
See Sec.~\ref{sec:simulations} for details on the simulations.}
\end{center}
\end{figure}
\begin{figure}[ht]
\begin{center}
\includegraphics[clip, width=0.7\linewidth]{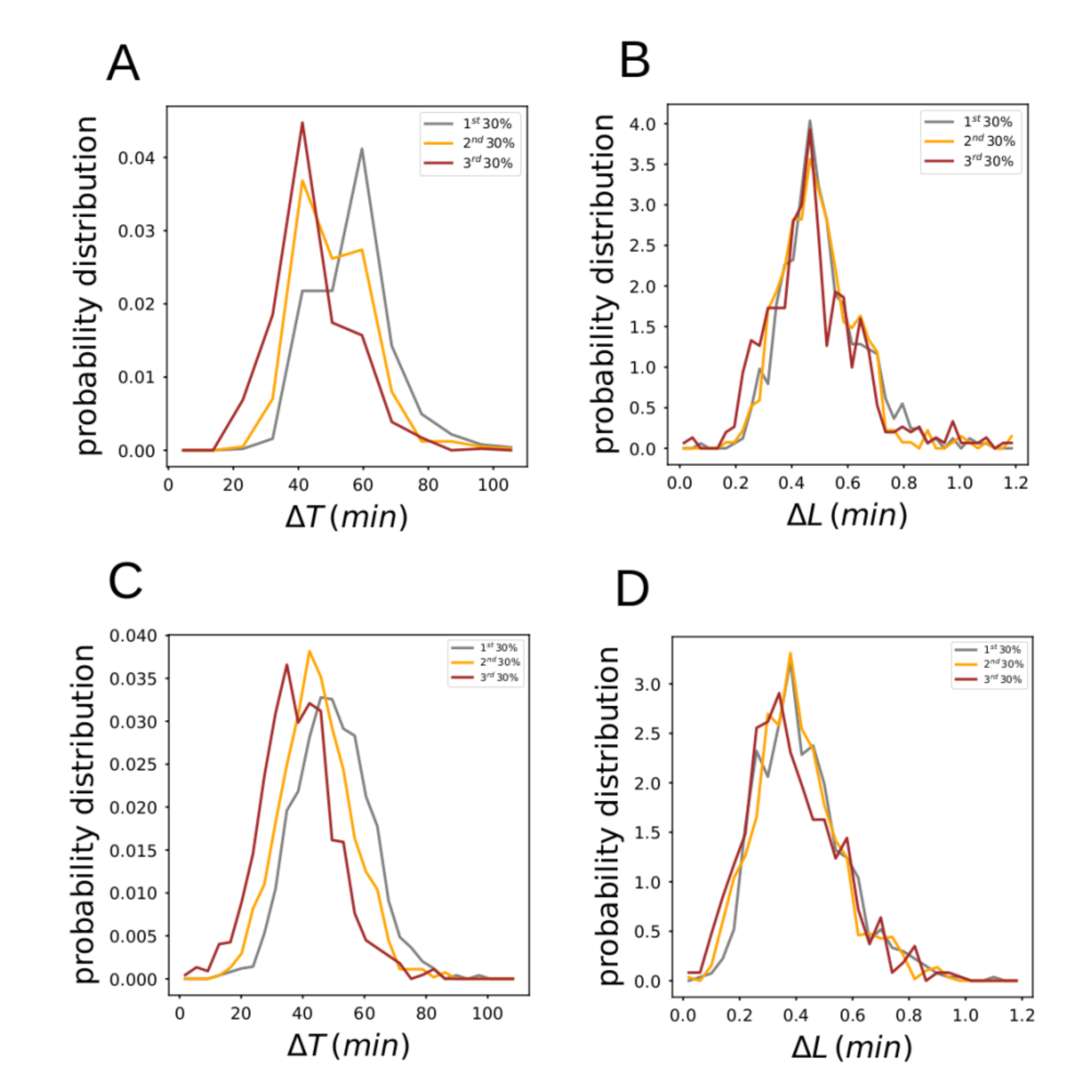}
\caption{\label{fig:conditional_distributions} Conditional distributions in experiments and model. (A)\&(B) The distributions of the division time ($\Delta T$) and the length elongation ($\Delta L$) when the size at birth $L_0$ is in three different percentiles from experimental measurements. The corresponding distributions from model results are shown in (C)\&(D). }
\end{center}
\end{figure}
In Fig.~\ref{fig:exp_vs_model}A, the division time $\Delta T$ is plotted as a function of the size at birth $L_0$ for every cell cycle (lighter dots), together with the mean corresponding to a specific division time (darker dots), for two nutrient conditions from our mother machine experiments. Consistent with previous experiments~\cite{Jun_Curr_Bio}, the size at birth has a negative correlation with respect to the division time, i.e., longer cells tend to divide faster. The mean value of the division time decreases for richer nutrient condition, suggesting that the rate of cell growth is influenced by the nutrient condition. 

In Fig.~\ref{fig:exp_vs_model}B, values of $\Delta L$ for each generation are shown as a function of $L_0$ (lighter dots). It is clear from the mean values (darker dots) that the elongation $\Delta L$ does not depend on the initial size $L_0$, but it slightly increases when the nutrient conditions are richer. The independence of $\Delta L$ on $L_0$ is also consistent with previous experiments~\cite{Jun_Curr_Bio}. In fact, this empirical observation was incorporated into the phenomenological \textit{adder model} \cite{Amir2014, Huang2017}, which starts from the assumption that a cell divides when its elongation reaches a fixed value independent of its size at birth $L_0$. The negative correlation between $\Delta T$ and $L_0$ and the independence of $\Delta L$ on $L_0$ are the two key features in cell size control we wish to reproduce from our growth-division model.

In Fig.~\ref{fig:exp_vs_model}C, We show the division time $\Delta T$ versus the size at birth $L_0$ for each individual cell cycle (lighter dots) from our model. We divided the values of $L_0$ into equally spaced bins and calculated the average of $\Delta T$ and $L_0$ for each bin (darker dots). The same analysis was used to plot the increment $\Delta L$ as a function of the size at birth $L_0$ for different cell cycles in Fig.~\ref{fig:exp_vs_model}D.
Our model reproduces the negative correlation between $L_0$ and $\Delta T$ as well as the independence of $\Delta L$ on $L_0$. Furthermore, the dependence of $\Delta T$, $L_0$, and $\Delta L$ on the nutrient conditions are also in quantitative agreement with our mother machine experiments.
Moreover, for a given nutrient condition, it is possible to appreciate the dependency of $\Delta T$ and $\Delta L$ on $L_0$ by considering the conditional probability distributions. In Fig.~\ref{fig:conditional_distributions} we plot the distributions of $\Delta T$ and $\Delta L$ for cell cycles in which the initial length $L_0$ was in specific ranges. We can thus notice that while $\Delta T$ tends to decrease for larger $L_0$, the distributions of $\Delta L$ collapse, and the results from the simulations are in line with the ones from the experiments.

\subsection{Derivation of the Langevin equation for the concentration using a mean field approximation}\label{sec:Langevin_c}
In this section and the next one we will derive the differential equation of the probability distribution $P(c)$ and the relation between the mean and the variance of the protein concentration in our mean field model.
To do so, we first write the time derivative of $c$ in terms of derivatives of $N$ and $L$:
\begin{equation}
    \frac{dc}{dt}=\frac{1}{L}\frac{dN}{dt}-\frac{c}{L}\frac{dL}{dt}\,.
\end{equation}
Using Eq.~\ref{eqn:main_eq} and Eq.~\ref{eqn:N} in the main text, the Langevin equation becomes:
\begin{equation}\label{eqn:conc_si}
   \frac{dc}{dt}=\hat{K}_n-\hat{K}_l\,c+c\sum_i^{n_d(t)}\left[\left(w_i^{(l)}-w_i^{(n)}\right)\delta(t-t_i)\right]\,.
\end{equation}
The two variables for the division term are
\begin{align}
w_i^{(l)}&=\frac{L_{a}}{L_{b}} \,,\\
w_i^{(n)}&=\frac{N_{a}}{N_{b}} 
\end{align}
where $L_{b}$, $N_{b}$, $L_{a}$ and $N_{a}$ are the values of the size and the protein number before and after the division respectively. 

The difference $\delta w_i=w_i^{(l)}-w_i^{(n)}$ has the following properties:
\begin{align}
    \left<\delta w_i\right> &=0\,,\nonumber\\
    \left<\delta w_i\delta w_j\right>&=2\Delta_w \delta_{ij}\,,
\end{align}
where $\Delta_w$ is the variance of the partition difference. 
We can thus define the noise due to cell division:
\begin{equation}
    \eta_d(t)\equiv \sum_i^{n_d(t)}\delta w_i\delta(t-t_i(Z))\,.
\end{equation}
This new stochastic variable has still mean equal to zero and it is delta-function correlated:
\begin{equation}
    \left<\eta_d(t)\right>=0\,,
\end{equation}
and
\begin{equation}\label{eqn:corr_etaD}
    \left<\eta_d(t_1)\eta_d(t_2)\right>=2\Delta_d \delta(t_2-t_1)\,,
\end{equation}
where 
\begin{equation}
    \Delta_d=\frac{1}{2\left<\Delta T\right>}\left<\left(\frac{L_{a}}{L_{b}}-\frac{N_{a}}{N_{b}}\right)^2\right>\,.    
\end{equation}
The other sources of noise are discussed in \ref{Sec:noise_strength}.

\subsection{Derivation of the square Taylor's law}\label{Sec:TL_derivation}
From the Langevin equation (Eq.~\ref{eqn:conc} in the main text), the steady state protein concentration distribution function $P(c)$ satisfies the stationary Fokker-Planck equation\footnote{We used the Ito interpretation for the multiplicative noise here. Using the Stratonovich interpretation does not change the qualitative results.}:
\begin{equation}\label{eqn:FP}
    \left<\hat{K}_l\right>\frac{d}{dc}\left[(\mu-c)P\right]=\frac{d^2}{dc^2}\left[D(c)P\right]\,,
\end{equation}
where $D(c)=\mu^2 \Delta_a + 2\mu c \Delta_{am} + c^2 \Delta_m$ with $\Delta_{(a,m,am)}$ representing the noise strength for $\eta_a$, $\eta_m$, and their correlation. The solution of Eq.~\ref{eqn:FP} is reported in the main text (Eq.~\ref{eqn:distribution}). Here we derive the relation between the mean and the variance. 

After a first integration over $c$ we obtain
\begin{equation}\label{eqn:after_int}
    \left<\hat{K}_l\right>\left[(\mu-c)P(c)\right]=\frac{d}{dc}\left[D(c)P(c)\right]\,.    
\end{equation}
If we integrate on both sides and we use the fact that $P(c)D(c)\to 0$ when $c\to\infty$ we have 
\begin{equation}
    \left<\hat{K}_l\right>(\mu -\left<c\right>)=0\,,
\end{equation}
and thus $\left<c\right>=\mu$.
If instead we multiply Eq.~\ref{eqn:after_int} on both sides for $(\mu-c)/\left<\hat{K}_l\right>$, and we integrate over $c$, on the left hand side we simply have $\sigma_c^2$. Therefore, after an integration by parts, the equation reduces to
\begin{equation}\label{eqn:eq_sigma_c}
    \sigma_c^2=\frac{1}{\left<\hat{K}_l\right>}\int D(c)P(c) dc=\frac{1}{\left<\hat{K}_l\right>}(\mu^2\Delta_a+2\mu^2\Delta_{am}+(\mu^2+\sigma_c^2)\Delta_m)\,.
\end{equation}
By solving Eq.~\ref{eqn:eq_sigma_c} we obtain the variance as a function of the noise strengths and $\mu$
\begin{equation}
    \sigma_c^2=\frac{\Delta_a+2\Delta_{am}+\Delta_m}{\left<\hat{K}_l\right>-\Delta_m}\mu^2= \frac{\Delta_d+\Delta_g}{\left<\hat{K}_l\right>-\Delta_m}\mu^2\, ,
    \label{TLE}
\end{equation}
where $\Delta_a =\left<\hat{K}_l\right>^2(\Delta_{(r)}+\Delta_n)$ is the noise strength for $\eta_a$, $\Delta_{am}=-\left<\hat{K}_l\right>^2\Delta_{(r)}$ is the correlation between $\eta_a$ and $\eta_m$, and $\Delta_m=\Delta_d+\left<\hat{K}_l\right>^2(\Delta_{(r)}+\Delta_l)$ is the strength of the noise $\eta_m$ with $\Delta_{(r)}=\langle \delta r^2\rangle/\langle r\rangle^2 $ the noise strength for $\delta r/\langle r\rangle $. The above equation (Eq.~\ref{TLE}) is the same as Eq.~10 in the main text.

\subsection{Detailed noise analysis}\label{Sec:noise_strength}
%{\color{red} Alberto, see the following for how to coarse grain noise in time. You can use this to derive the noise strength in the Langevin equation from the full model with its parameters; as well as from the correlation function computed from the experimental data.  {\bf How to coarse-graining noise in time:} Given a random variable $\eta(t)$ with zero mean, i.e., $\langle \eta(t)\rangle =0$ and correlation $\langle \eta(t_1) \eta(t_1+t)\rangle_{t_1} =C(t)$, where $C(t)$ is an even function of $t$ and it decays with time. If we consider the noise at a time scale that is much larger than the correlation time, we can approximate the noise $\eta$ by its coarse-grained form $\tilde{\eta}(t)$ with zero mean and a delta-function correlation: $\langle \tilde{\eta}(t_1)\tilde{\eta}(t_1+t)\rangle_{t_1}=2\Delta \delta(t).$ The connection between the strength of the coarse-grained noise and the correlation for the original noise can be obtained by requiring the integrated correlation over time to be the same: $2\Delta =\int_{-\infty}^{\infty} C(t)dt$, which lead to: $$\Delta =\int_0^\infty C(t)dt.$$ For example, if the correlation function is an exponential function: $C(t)=C(0)\exp(-t/\tau)$ with a correlation time $\tau$, we have: $\Delta =C(0)\tau$. Note that $C(0)$ is just the variance of the noise $\eta$. }

The averages of the effective growth and expression rates $\left<\hat{K}_{l,n}\right>$ can be calculated from experiments in the following way:

\begin{align}
    \left<\hat{K}_l\right>&=\left<\frac{1}{L}\frac{\delta L}{\delta t}\right>\,,\nonumber\\
    \left<\hat{K}_n\right>&=\left<\frac{1}{L}\frac{\delta N}{\delta t}\right>\,,   
\end{align}
where $\delta L$, $\delta N$ and $\delta t$ are the smallest increment allowed by the experimental set up. In our case, the time step was $\delta t=5\, \mbox{min}$. From the experiments we do not have direct information about $R$ and the noise from the observable variables includes contributions from $r$ and contributions from $N$ and $L$. In other words, we can directly measure $\chi_l$ and $\chi_n$ but not $\eta_l$ and $\eta_n$. 
The noise strengths of  $\chi_l$ and $\chi_n$ are  given by the following formula:
\begin{equation}
    \Delta_{(\chi_l, \chi_n)}=\frac{1}{2}\int ds\, 
    \left<\chi_{(l,n)}(t)\chi_{(l,n)}(t+s)\right>\,.
\end{equation}
Here the average is over an ensemble of time traces and the result will not depend on the specific time $t$. 

Since the noises $\eta_n$ and $\eta_l$ are independent, when we calculate the correlation between $\chi_l$ and $\chi_n$ we  obtain the strength of the noise $\delta r/\left<r\right>$ -the only non zero term in the correlation- that we call $\Delta_{(r)}$ (not to be confused with the strength of $\eta_r$) :
\begin{equation}
    \Delta_{(r)}=\frac{1}{2}\int\,ds\,\left<\chi_l(t)\chi_n(t+s)\right> \,.
\end{equation}
Once we have $\Delta_{(r)}$, by subtracting it from $\Delta_{\chi_l}$ and $\Delta_{\chi_n}$ we will obtain $\Delta_l$ and $\Delta_n$ respectively.  

\begin{table}
\begin{tabular}{ |c|c|c|c|c|c|c|c|c| }
 \hline
conditions & $\Delta_{\chi_l}$ (min) & $\Delta_{\chi_n}$ (min) & $\left<K_l\right>\, (\mbox{min}^{-1})$ & $\Delta_{(r)}$ (min) & $\Delta_m\, (\mbox{min}^{-1})$ & $\Delta_d\, (\mbox{min}^{-1})$ & $\Delta_g/\Delta_d$ & $A$ \\
 \hline
%Rich A & 0.506 & 2.4 & 0.017 & 0.59 & 0.0002 & 4.5e-05 & 12 & 0.034 \\
Rich 4 & 0.619 & 2.4 & 0.017 & 0.66 & 0.00023 & 5.5e-05 & 8.7 & 0.032 \\
Rich 5 & 0.412 & 2.8 & 0.017 & 0.12 & 0.00015 & 3.7e-05 & 22 & 0.053 \\
Rich B & 0.429 & 1.7 & 0.017 & 0.56 & 0.00017 & 4.5e-05 & 6.8 & 0.021 \\
Rich C & 0.654 & 2.3 & 0.016 & 0.63 & 0.0002 & 3.7e-05 & 12 & 0.03 \\
Rich D & 0.904 & 1.8 & 0.015 & 0.85 & 0.00022 & 3e-05 & 6.9 & 0.017 \\
%Poor A & 8.8 & 8.6 & 0.0093 & 1.6 & 0.00083 & 6.1e-05 & 20 & 0.15 \\
Poor 4 & 9.42 & 8.5 & 0.0075 & 5.3 & 0.00057 & 3.9e-05 & 11 & 0.065 \\
Poor 5 & 8.2 & 6.25 & 0.0092 & 4.3 & 0.00074 & 3.9e-05 & 12 & 0.063 \\
Poor B & 8.2 & 5.15 & 0.0087 & 3.5 & 0.00063 & 2.5e-05 & 22 & 0.061 \\
Poor C & 4.7 & 4.1 & 0.0092 & 2.2 & 0.00042 & 2.1e-05 & 18 & 0.045 \\
Poor D & 5.49 & 6.3 & 0.0087 & 3.4 & 0.00044 & 1.9e-05 & 19 & 0.048 \\
%Poor 1 & 8.26 & 13 & 0.0078 & 3 & 0.00054 & 4.1e-05 & 22 & 0.13 \\

 \hline
\end{tabular}
\caption{\label{table:parameters}Relevant parameters inferred from the experiments for different promoter strengths and nutrient conditions. These values have been calculated by means of averages over different mother cells.}
\end{table}

\begin{figure}[ht]
\begin{center}
\includegraphics[clip, width=0.9\linewidth]{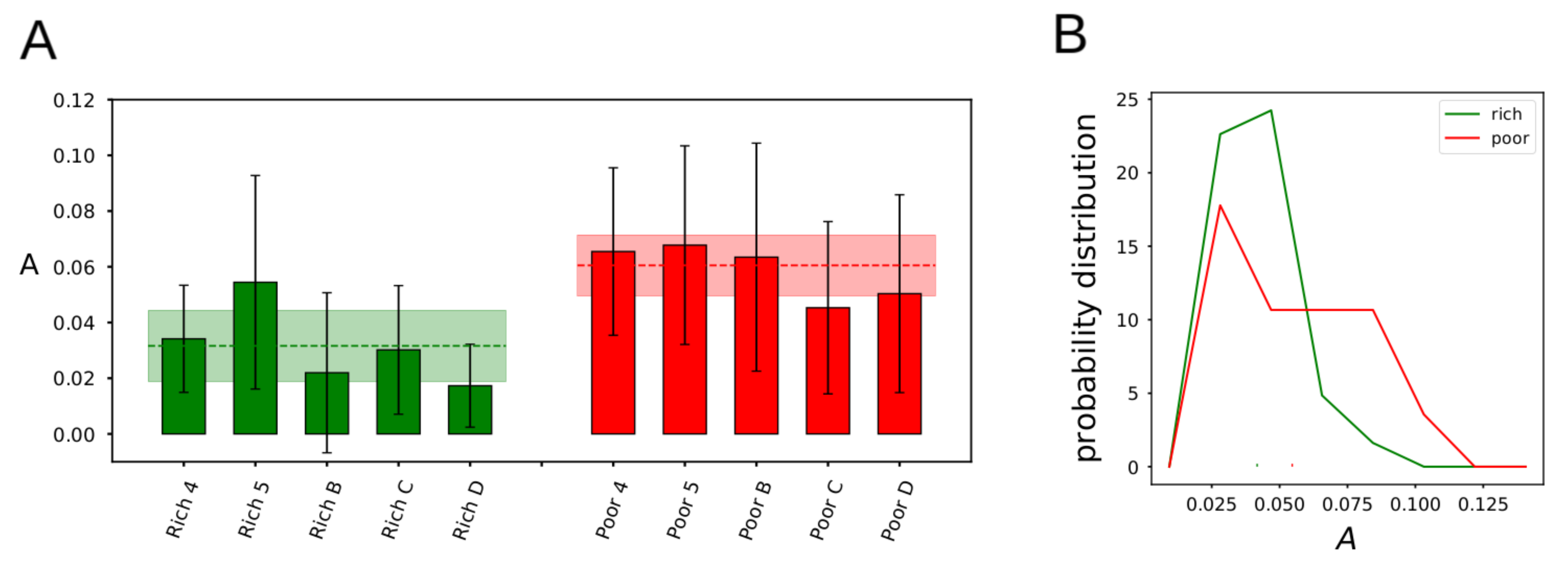}
\caption{\label{fig:A_bar_chart}(A) Scaling prefactor constant $A$ averaged over all mother cells within a population with different combinations of nutrient conditions and promoter strengths. The dashed line indicates the average for a specific nutrient condition. The colored region indicates the standard deviation of the mean for each nutrient condition.
(B) Distribution of $A$ from individual mother cells in rich and poor media.}
\end{center}
\end{figure}

We can now write $\Delta_a$, $\Delta_{am}$ and $\Delta_m$ in terms of these noise strengths:
\begin{align}
    \Delta_a &=\left<\hat{K}_l\right>^2\Delta_{\chi_n}\,,\nonumber\\
    \Delta_m &=\Delta_d+\left<\hat{K}_l\right>^2\Delta_{\chi_l}\,,\nonumber\\
    \Delta_{am}&= -\left<\hat{K}_l\right>^2\Delta_{(r)} \nonumber
\end{align}
Given these expressions, the parameter $A$ can be written as 
\begin{equation}\label{eqn:Coef}
    A=\frac{\Delta_g+\Delta_d}{\left<\hat{K}_l\right>-\Delta_m}\,,
\end{equation}
where we have defined the \textit{growth} noise strength:
\begin{equation}\label{eqn:growth_strength}
    \Delta_g=\left<\hat{K}_l\right>^2(\Delta_{\chi_n}+\Delta_{\chi_l}-2\Delta_{(r)})
\end{equation}
In Fig.~\ref{fig:A_bar_chart}A we show the average coefficient $A$ for all the different conditions and promoter strengths. In Fig.~\ref{fig:A_bar_chart}B the distributions of $A$ are shown for the two nutrient conditions, sampling together the values of the mother cells for all promoter strengths. 
The noise strengths and other relevant parameters  relative to all combinations of nutrient conditions and promoter strengths are reported in Table~\ref{table:parameters}.

\begin{figure}[ht]
\begin{center}
\includegraphics[clip, width=0.9\linewidth]{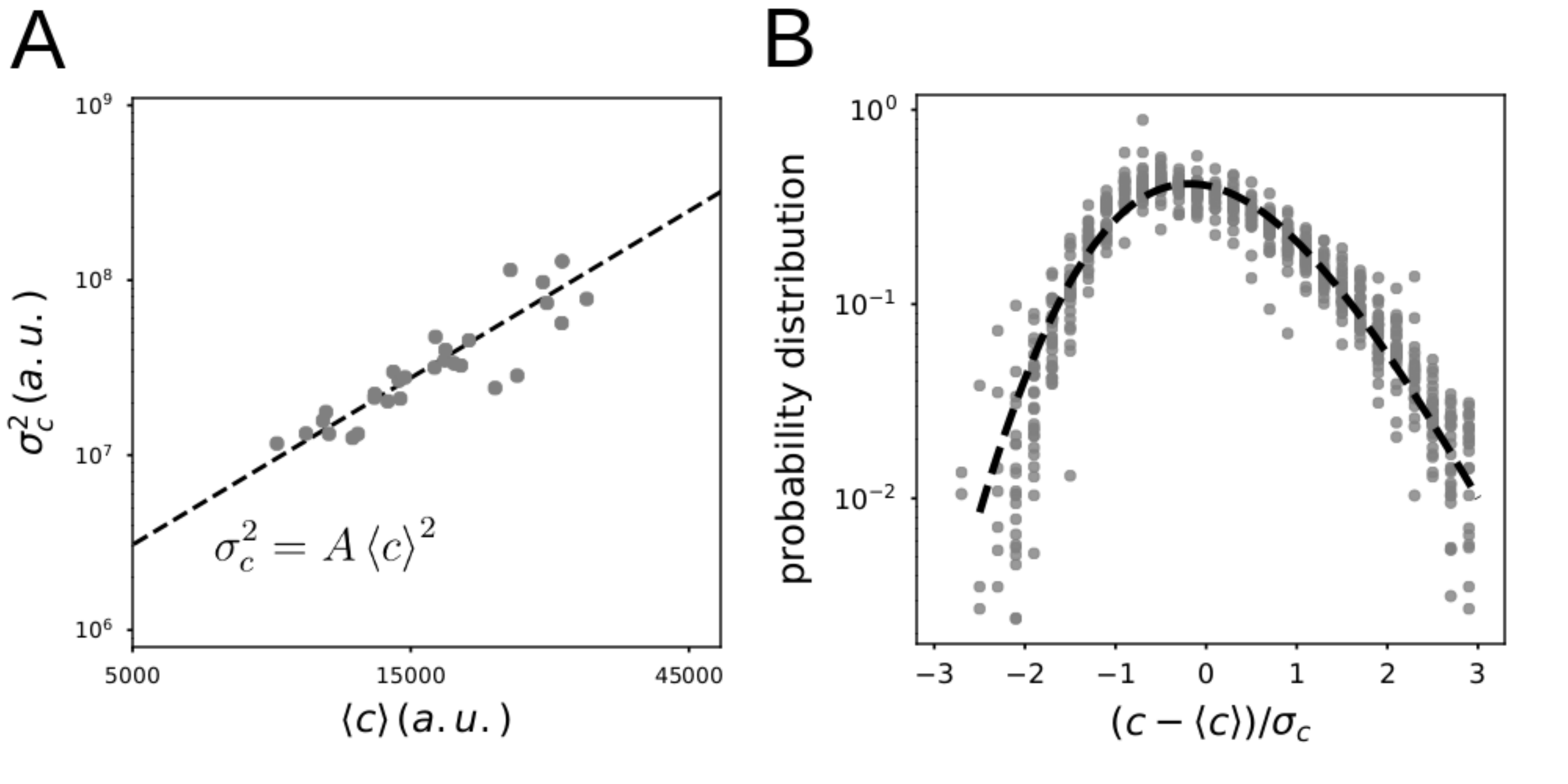}
\caption{\label{fig:Salman_distribution} 
(A) Mean of the fluorescence density $c$ as a function of the variance from the experimental data obtained in \cite{Salman2012}. The dashed line is obtained using an average value for the coefficient $A=\sigma_c^2/\left<c\right>^2$. The value obtained in this set of data was $\left<A\right>=0.12$. 
(B) The scaled protein concentration distributions from the experimental data reported in ~\cite{Salman2012}. The black curve is the distribution that we show in the main text. 
%The black curve represents an inverse Gamma distribution based on the mean and variance of the experimental data. The distribution obtained in the main text reduces asymptotically to the Gamma distribution whenever the contribution to the noise from the multiplicative term ($\Delta_m$) is dominant with respect to the one from the additive term ($\Delta_a$). 
 }
\end{center}
\end{figure}

\newpage
\bibliography{biblio_cell_div}{}
\bibliographystyle{unsrt}

\end{document}